\documentclass[aps,prd,preprintnumbers,groupedaddress,nofootinbib,amssymb,eqsecnum,notitlepage]{revtex4}
\usepackage{amsfonts}
\pdfoutput=1
\usepackage{graphicx}  
\usepackage{dcolumn}   
\usepackage{bm}        
\usepackage{amssymb}   
\usepackage{amsmath}
\usepackage{mathtools}
\usepackage{cases}
\usepackage{color}
\usepackage{epsfig}
\usepackage[normalem]{ulem}
\usepackage{amssymb,amsmath,bm,bbm}
\usepackage{verbatim}
\usepackage{appendix}
\newcommand{\mpl}{M_{\text{Pl}}}

\newcommand{\be}{\begin{equation}}
\newcommand{\ee}{\end{equation}}
\newcommand{\bea}{\begin{eqnarray}}
\newcommand{\eea}{\end{eqnarray}}

\begin{document}
\pdfoutput=1

\begin{flushright}
    YITP-19-27
\end{flushright}

\title{Primordial Tensor Perturbation in Double \\Inflationary Scenario with a Break}

\author{Shi Pi$^{1}$,
Misao Sasaki$^{1,2,3}$ and
Ying-li Zhang$^{4,5}$}

\affiliation{$^1$Kavli Institute for the Physics and Mathematics of the Universe (WPI), Chiba 277-8583, Japan\\
$^2$Yukawa Institute for Theoretical Physics, Kyoto University, Kyoto 606-8502, Japan\\
$^3$Leung Center for Cosmology and Particle Astrophysics, National Taiwan University, Taipei 10617\\
$^4$Department of Physics, Tokyo Institute of Technology, 2-12-1 Ookayama, Meguro-ku, Tokyo 152-8551, Japan\\
$^5$Department of Physics, Faculty of Science, Tokyo University of Science, 1-3, Kagurazaka,
Shinjuku-ku, Tokyo 162-8601, Japan}

\date{\today}

\begin{abstract}
We study the primordial tensor perturbation produced from the double inflationary scenario with an intermediate break stage. 
Because of the transitions, the power spectrum deviates from the vacuum one and there will
appear oscillatory behavior.
In the case of a scalar-type curvature perturbation, it is known that the amplitude of these
oscillations may be enhanced to result in the power spectrum larger than the one for the vacuum
case. One might expect the similar enhancement for the tensor perturbation as well.
Unfortunately, it is found that when the equation of state (EOS) parameter $w=p/\rho$ of the
 break stage is a constant with $w>-1/3$, 
the amplitude of oscillations is never large enough to enhance the power spectrum.
On the contrary,  the power spectrum is found to be suppressed even on those scales that
leave the horizon at the first inflationary stage and remain superhorizon throughout the entire
stage.  We identify the cause of this suppression with the correction terms in additional to the leading order constant solution on superhorizon scales. 
We argue that our result is general in the sense that any intermediate break stage
during inflation cannot yield an enhancement of the tensor spectrum
 as long as the Hubble expansion rate is non-increasing in time.
\end{abstract}


\maketitle

\section{Introduction}	
Inflation is a successful paradigm to resolve the horizon, flatness and unwanted relics problems 
in the Hot Big Bang cosmology. It also provides the initial seeds for fluctuations we observe today
in the cosmic microwave background (CMB) and in the large-scale structure. 
There have been a number of observations
aiming to measure and constrain the cosmological parameters,
to mention a few among others are
Wilkinson Microwave Anisotropy Probe (WMAP)~\cite{WMAP:2012}, Planck~\cite{Planck2018} 
and Sloan Digital Sky Survey (SDSS)~\cite{SDSS:2006}. \\ 

One of the most important predictions of inflation is generation of primordial gravitational waves (GWs).
The primordial GWs originate from the vacuum tensor fluctuations during inflation, which are stretched
out of the horizon resulting in an almost scale-invariant spectrum.
After inflation, they re-enter the horizon at different epochs, and are redshifted until today~\cite{Guzzetti:2016mkm,Cai:2017cbj}. 
The long wavelength modes which re-enter the horizon in the matter dominated era are less redshifted, 
and may be detected in the B-mode polarization of the CMB anisotropy. 
Unfortunately the current observation by Planck and BICEP/Keck only gives the upper bound on the tensor-to-scalar ratio, $r\equiv \mathcal{P}_T/\mathcal{P}_{S}<0.07$ at 95$\%$ confidence~\cite{BICEP2:2015}.
It is hoped that he future experiments like AliCPT~\cite{Li:2017drr} or 
LiteBIRD~\cite{Matsumura:2013aja}  may detect the tensor perturbation.\\

Primordial GWs with shorter wavelengths are expected to give
an almost scale-invariant power spectrum of order $\Omega_\text{GW}\sim10^{-16}(r/0.1)$ at 
frequency $\gtrsim10^{-14}$Hz. 
Pulsar timing array~\cite{Hobbs:2009yy,Carilli:2004nx} or ground/space based laser 
intefereometers like LIGO-VIRGO-KAGRA~\cite{Aasi:2013wya}, ET~\cite{Punturo:2010zz,Sathyaprakash:2012jk},
LISA~\cite{AmaroSeoane:2012km,AmaroSeoane:2012je,Audley:2017drz}, 
Taiji~\cite{Guo:2018npi}, Tianqin~\cite{Luo:2015ght}, BBO~\cite{Crowder:2005nr,Corbin:2005ny} 
or DECIGO~\cite{Kawamura:2006up,Kawamura:2011zz} 
are aimed to detect the GWs with such higher frequencies. 
However, a small tensor-to-scalar ratio indicates that it is difficult to 
see the primordial GWs by most of these detectors, 
if the amplitude of the primordial tensor perturbation on small scales is 
at most the same as that on the CMB scales. 
Although it is still possible that the primordial tensor perturbation has an observable 
amplitude on small scales, according to the current observational constraints on the 
spectral tilt and its running~\cite{Green:2018akb}, to realize such a blue-tilted power 
spectrum for the tensor perturbation is not an easy task~\cite{Cai:2014uka}.\\

On the other hand, the amplification of the curvature perturbation is much easier. 
For instance, the multiple inflationary scenario which includes two or more inflationary periods
 intervened by a stage of decelerated expansion could have  characteristic signatures on certain scales.
 Especially, the double inflationary scenario, or inflation with an intermediate break, 
 was originally introduced to decouple the power spectrum on small scales from large (CMB)
  scales~\cite{Kofman:1985, Silk:1987, Zelnikov:1991JETP, Polarski:1992, Polarski:1995, Adams:1997}.\\
   
 In the framework of supersymmetric particle physics, various multiple inflation models have been
  proposed~\cite{Kanazawa:2000, Lesgourgues:2000, Yamaguchi:2001, Yamaguchi:2001a, Burgess:2005, Kawasaki:2011,Yamaguchi:2011,Maeda:2018sje}. 
  One of the typical features of double inflation models is the temporal violation of the 
  standard slow-roll condition during the transition between two stages of inflation, 
  which could lead to an enhancement of the primordial curvature perturbation.
  Recently this scenario received a lot of attention,
  since such an amplification of the scalar perturbation could seed the formation of primordial 
  black holes
  (PBHs)~\cite{Bellido:1996,Kanazawa:1998,Kanazawa:2000a,Kawasaki:2006,Clesse:2015,
  	Kawasaki:2016,Inomata:2017,Inomata:2018,Pi:2018,Cai:2018} (for a review, see e.g.\cite{Sasaki:2018}) 
  and induce observationally testable GWs~\cite{Ananda:2006af,Baumann:2007zm,Osano:2006ew,Alabidi:2012ex,Alabidi:2013wtp,Inomata:2016rbd,Orlofsky:2016vbd,Kohri:2018awv,Assadullahi:2009jc,Biagetti:2014asa,Nakama:2016gzw,Gong:2017qlj,Giovannini:2010tk,Garcia-Bellido:2017aan,Cai:2018dig,Inomata:2018epa,Unal:2018yaa,Byrnes:2018txb,Cai:2019jah}.
  For example, in~\cite{Pi:2018}, a double inflation model was studied in which there appears an
  oscillatory behavior in the curvature perturbation power spectrum
  due to damped oscillations of the inflaton at the end of the first stage, and it was found
  that the first peak of the damped oscillations
  may produce a large enhancement that leads to the PBH formation.\\

Inspired by these results, we consider if there is a similar enhancement mechanism for
 the primordial tensor perturbation at around the intermediate break of a double inflation model. 
Even without an oscillatory behavior in the background dynamics, one typically expects
the appearance of an oscillatory feature in the power spectrum due to a phase transition 
that causes excitations from the vacuum state, as in the case of the scalar-type curvature
perturbation.
Thus it is of interest to clarify if such an oscillatory feature in the spectrum could 
become large enough to enhance the amplitude also in the case of the tensor perturbation.\\

 To study the effect of an intermediate break stage, we consider instotaneous phase transitions
both at the end of the first inflationary stage and the beginning of the second
inflationary stage, and model the break stage as that of a constant equation
 of state parameter, that is, a stage with $w=p/\rho=const$. We also assume 
 exact de Sitter background for both the first and second stages of inflation for simplicity.
 These assumptions make us possible to perform an exact analytical study,
 albeit that it involves special functions.
 In particular, we make detailed analyses in the case of $w=1/3$ (radiation-domination) and 
 $w=0$ (matter-domination).
 Since it is naturally expected that a sharper transition will produce a larger enhancement,
 an instantaneous phase transition we consider may be regarded as the limiting case 
 where one would obtain a maximum possible enhancement, if at all.\\

Specifically, we set the initial state to be the natural vacuum state deep inside the horizon,
and compute the evolution of the mode function by matching it at the two transition
epochs. We then give an analytical formula of the tensor power spectrum 
for double inflation with an intermediate stage EOS $w>-1/3$.
We then explicitly evaluate it in the case of $w=1/3$ and $0$. 
 Unfortunately, although there appear oscillations in the spectrum,
 the amplitude is found to be not large enough to give any enhancement.\\

On the contrary, we find that the spectrum is actually suppressed relative to the original 
vacuum spectrum already at wavelengths that exit the horizon at the first inflationary stage and
that are long enough so that they never re-enter the horizon during the entire stage 
 (see Fig.~\ref{fig:exact} in the range of $p\equiv k/k_1<1$). 
This behavior is caused by the correction terms of order $z^2\equiv (k\eta)^2$
to the constant mode on superhorizon scales, together with the effect of the decaying mode. 
To clarify this suppression in a physically more intuitive way, we employed an approximation
in which the mode function on subhorizon scales is given by the WKB approximation
and the constant mode on superhorizon scales is corrected by the $O(z^2)$ term.
The power spectra obtained from the exact solutions and the approximate ones 
agree with each other very well (see Fig.\ref{fig:compare} in the range of $p<1$).
This confirms that the superhorizon suppression of the power spectrum for 
those long wavelength modes is due to the $O(z^2)$ correction term.\\

This paper is organized as follows: In Sec.\ref{sec:background}, we formulate our model
of double inflation with an intermediate break stage and introduce a set of convenient parameter
 that characterize it. 
In Sec.\ref{sec:exact}, we derive the exact solutions for the tensor mode functions by solving 
the equations of motion (EOM) and matching conditions at the two boundary epochs of inflation. 
Then an analytical formula for the power spectrum is derived. 
We then explicitly evaluate the power spectrum for $w=1/3$ and $0$ cases.
We find no enhancement. On the contrary, we find an appreciable suppression on large scales.
In Sec.\ref{WKBsec}, we consider an approximation in which the
 correction term proportional to $z^2$ on superhorizon scales is taken into account,
and derive the power spectrum for general $w>-1/3$. We find that the formula agrees 
with the exact analytical result expanded up through the $O(z^2)$ corrections.
In Sec.~\ref{conclude}, we conclude our results. 
 In Appendix \ref{app:a}, we give some basic formulas of the scale factor in terms of 
 the matter EOS used in the text.
  In Appendix \ref{inteconstant}, we present the relations among the 
 integration constants in the expressions for the scale factor in different stages,
 as well as useful formulas derived from those relations.
 In Appendix \ref{Normalization}, we present useful expressions obtained from the
 normalization condition of the mode functions.
 In Appendix \ref{WKBsuppress}, we present a detailed
derivation of the approximate solutions with the $O(z^2)$ correction terms.
 In Appendix \ref{nogo}, we prove a no-go theorem for enhancement of the amplitude of
 the tensor power spectrum, by showing that the suppression on large scales 
 is inevitable for any EOS with $w>-1/3$ between the two inflationary stages.

\section{Background spacetime}	\label{sec:background}

We first specify the background spacetime. We assume a spatially flat expanding universe,
\begin{align}
ds^2&=-dt^2+a^2(t)\delta_{ij}dx^idx^j
\nonumber\\
&=a^2(\eta)(-d\eta^2+\delta_{ij}dx^idx^j)\,,
\end{align}
where the scale factor is regarded as a function of either the cosmic time $t$ or the
conformal time $\eta$ interchangeably. We assume there are two stages of inflation
with an intermediate break stage. We assume that the Hubble parameter 
$H\equiv\dot{a}/a$ ($\dot{~}=d/dt$) is constant at both inflationary stages, and
the EOS parameter $w\equiv p/\rho$ is a constant satisfying $w>-1/3$ during the break stage.
We denote the Hubble parameters at the first stage by $H_1$ and the second stage by
$H_2$, respectively, and the first transition time by $t_1$ ($\eta_1$) and the second by
$t_2$ ($\eta_2$). 
 Then the scale factor $a(\eta)$ is expressed in terms of the conformal time as
\be\label{a-eta}
a(\eta)=\left\{\begin{matrix}
\displaystyle a_{\mathrm{I}}=-\frac1{H_1(\eta-\gamma_1)}\,; & \eta<\eta_1\,,&~\\
\\
\displaystyle a_{\mathrm{II}}=\alpha\left(\eta-\beta\right)^n\,\quad; & \eta_1<\eta<\eta_2
\quad& \left(n=\dfrac{2}{1+3w}\right)\,,\\
\\
\displaystyle a_{\mathrm{III}}=-\frac1{H_2(\eta-\gamma_2)}\,;  & \eta_2<\eta\,.&~
\end{matrix}
\right.
\ee
where $0<n<\infty$ for $w>-1/3$, and
$\gamma_1$, $\gamma_2$, $\alpha$ and $\beta$ are four integration constants. 
These constants are determined by the matching conditions that the scale 
factor $a(\eta)$ is $\mathbf{C}^1$-continuous at both $\eta=\eta_1$ and $\eta_2$. 
The details are given in Appendix \ref{inteconstant}. Here we quote the result:
\begin{align}
\gamma_1&=\eta_1+\frac{\Delta\eta}{n(s-1)}\,,\qquad \gamma_2=\eta_2+\frac{s\Delta\eta}{n(s-1)}\,,\\
\alpha&=\frac{n}{H_1}\left(\frac{s-1}{\Delta\eta}\right)^{n+1}\,,\qquad \beta=\eta_1-\frac{\Delta\eta}{s-1}\,,
\end{align}
where $\Delta\eta\equiv\eta_2-\eta_1$ and $s\equiv (H_1/H_2)^{1/(n+1)}$. 

\begin{figure}
\centering
\includegraphics[width=.7\textwidth]{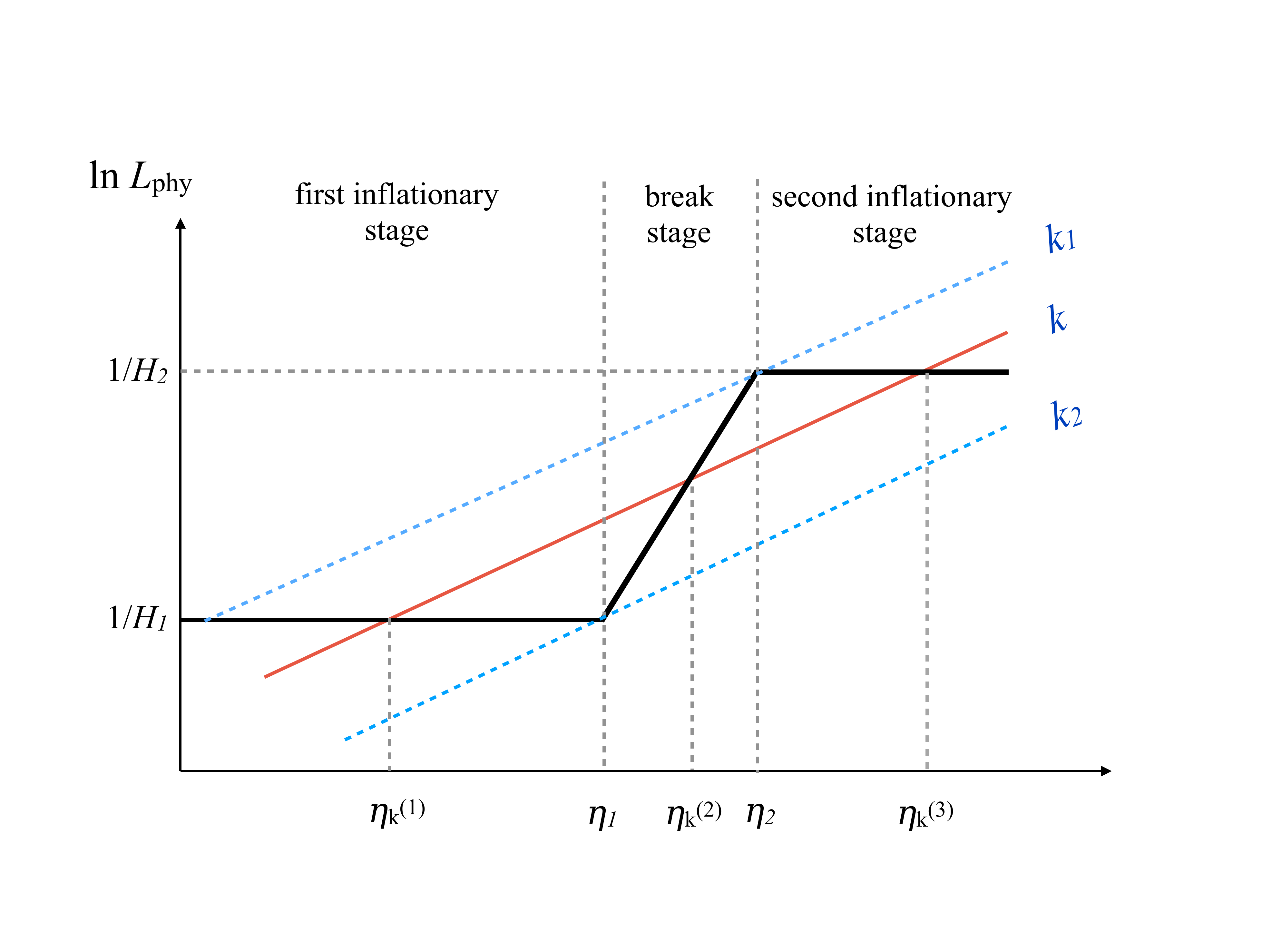}
\caption{Schematic diagram to demonstrate the the boundaries of wavenumber and the corresponding junction points in terms of conformal time $\eta$. The qualities $k_1$ and $k_2$ are defined in \eqref{k1} and \eqref{k2}, while $\eta^{(1)}_k$, $\eta^{(2)}_k$ and $\eta^{(3)}_k$ are defined in \eqref{etaI}--\eqref{etaII}, respectively. The modes with wavenumber $k<k_1$  correspond to the long wavelength modes which exit the horizon before $\eta^{(1)}_k$ and never re-enter the horizon until the second inflationary stage ends. Those modes with wavenumber $k>k_2$ correspond to the short wavelength modes which never exit the horizon during the whole double-inflationary epoch. The intermediate modes with wavenumber $k_1<k<k_2$ (remarked by red straight line) are the ones we are mostly interested in. It exits the horizon at $\eta=\eta^{(1)}_k<\eta_1$ during the first stage of inflation with Hubble parameter $H_1$, then re-enters the horizon at $\eta^{(2)}_k<\eta_2$ during the ``break'' stage when the scale factor $a(\eta)=\alpha(\eta-\beta)^n$.  After that, it evolves inside the horizon until $\eta^{(3)}_k$, then again exits the horizon in the second inflationary stage.}
\label{fig:sppic}
\end{figure}

The schematic diagram of space-time is shown in Fig.~\ref{fig:sppic}. 
The evolution of the background is divided into three stages: the first inflationary stage, 
the break stage, and the second inflationary stage.
There are two special values of the comoving wavenumbers $k_1$ and $k_2$, 
which are expressed as
\bea
k_1&=&\mathcal{H}(\eta_2)=-\frac1{\eta_2-\gamma_2}=\frac{n(s-1)}{s\Delta\eta},\label{k1}\\
k_2&=&\mathcal{H}(\eta_1)=-\frac1{\eta_1-\gamma_1}=\frac{n(s-1)}{\Delta\eta}=sk_1\,,\label{k2}
\eea
where $\mathcal H\equiv a'/a$ (${}'=d/d\eta$). 
The modes $k<k_1$ are those which never re-enter the horizon during the entire stage of
inflation once they exit the horizon at the first stage. 
The modes $k_1<k<k_2$ (represented by a red straight line) exit the horizon 
at $\eta=\eta^{(1)}_k<\eta_1$ during the first stage of inflation, re-enter the horizon at $\eta=\eta^{(2)}_k<\eta_2$ during the break stage, and exist the horizon again at $\eta=\eta^{(3)}_k$
during the second inflationary stage. 
The epochs $\eta=\eta_k^{(1)}$, $\eta_k^{(2)}$ and $\eta_k^{(3)}$ characterize 
the horizon crossings of the mode $k$, determined by $k={\cal H}(\eta)$,
and are given by
\begin{gather}\label{etaI}
\eta_k^{(1)}=\gamma_1-\frac1k,\\\label{etar}
\eta_k^{(2)}=\beta+\frac{n}{k},\\\label{etaII}
\eta_k^{(3)}=\gamma_2-\frac1k.
\end{gather}

\section{tensor mode functions}\label{sec:exact}

Let us consider the tensor perturbation $h_{ij}$ $(i,j=1,2,3)$ on our background,
\be\label{metric}
ds^2=a^2(\eta)\left[-d\eta^2+\left(\delta_{ij}+h_{ij}\right)dx^idx^j\right]\,,
\ee
where $\partial^i h_{ij}=h^i_{~i}=0$.
Inserting the above metric into the Einstein-Hilbert action, the second-order action 
for $h_{ij}$ is obtained as
\begin{align}
S^{(2)}&=\frac{\mpl^2}{8}\int dt\,d^3x\,
a^3(t)\left(\left(\dot{h}_{ij}\right)^2-\frac{1}{a^2}\left(\nabla h_{ij}\right)^2\right)
\nonumber\\
&=\frac{\mpl^2}{8}\int d\eta\,d^3x\,
a^2(\eta)\left(\left({h}_{ij}'\right)^2-\left(\nabla h_{ij}\right)^2\right)\,,
\label{Sori}
\end{align}
where $\mpl\equiv1/\sqrt{8\pi G}$. 
As usual, we expand $h_{ij}$ in Fourier modes and perform the Fox quantization,
\be
h_{ij}(\eta,\mathbf{x})=
\int \frac{d^3k}{(2\pi)^{3/2}}\sum_{\lambda=+,\times}
e_{ij}^{(\lambda)}({\bf k})\left(a_{\lambda,{\bf k}}h_{k}(\eta)e^{i{\bf k}\cdot{\bf x}} + h.c.\right)
\ee
where $e_{ij}^{(\lambda)}(\mathbf{k})$ is the polarization tensor satisfying
$e^{(\lambda)}_{ij}=e^{(\lambda)}_{ji}$, $\delta^{ij}e_{ij}^{(\lambda)}=e^{(\lambda)}_{ij}k^j=0$, 
$\delta^{ik}\delta^{j\ell}e^{(\lambda)}_{ij}e^{(\lambda')}_{k\ell}=\delta^{\lambda,\lambda'}$,
with $\lambda=+$ and $\times$ being 
the two independent GW polarization states,
$a_{\lambda,{\bf k}}$ is the annihilation operator, and
$h_k(\eta)$ is the positive frequency mode function associated with the vacuum
defined by $a_{\lambda,{\bf k}}|0\rangle=0$. 
The EOM for the mode function is given by
\be\label{eomh}
h_{k}^{\prime\prime}+2{\cal H}h_k' +k^2 h_k=0,
\ee
with the Klein-Gordon normalization condition,
\begin{align}
\left(\frac{M_{Pl}}{2}\right)^2\left(h_k\bar{h}_k'-\bar{h}_kh_k'\right)=\frac{i}{a^2}\,.
\label{KGnorm}
\end{align}
where $\bar{h}_k$ denotes the complex conjugate.
The  equation \eqref{eomh} can be solved analytically by Bessel functions,
which we perform below.

\subsection{Solution for $\eta<\eta_1$ }

At the first stage of inflation, $\eta<\eta_1$, we take the standard vacuum where
the positive frequency mode function approaches that of the Minkowski one proportional 
to $e^{-ik\eta}$ deep inside the horizon. The solution is given by 
\begin{align}\label{h1sol}
h_{\mathrm{I}}(\eta)=\frac{\sqrt{2}H_1}{\mpl k^{3/2}}\left(z_1+i\right)e^{iz_1}
\end{align}
where $z_1\equiv-k(\eta-\gamma_1)$, and we have omitted the mode index $k$
while attaching the subscript $\mathrm{I}$ to denote that it is the mode function at the first stage. 

\subsection{Matching at $\eta=\eta_1$}
At the stage $\eta_1<\eta<\eta_2$, the scale factor is proportional to $(\eta-\beta)^n$
$\bigl(n=1/(1+3w)\bigr)$. Then the general solution for the mode function $h_{\mathrm{II}}$ 
can be expressed as 
\begin{align}\label{h2exact}
h_{\mathrm{II}}=\frac{\sqrt{2\pi}}{\alpha\mpl}\left(\frac{z_2}{k}\right)^{-\nu}
\left[C_1J_{\nu}(z_2)-iC_2Y_{\nu}(z_2)\right]\,,
\end{align}
where $z_2\equiv k(\eta-\beta)$, $\nu\equiv n-1/2$, $J_{\nu}(z_2)$ and $Y_{\nu}(z_2)$ 
are the Bessel functions of the first and second kinds, respectively, and $C_1$ and $C_2$ 
are the two constants to be determined by the matching conditions. 
Detailed derivation is given in Appendix~\ref{Normalization}.
Note that the Klien-Gordon normalization gives the condition,
\begin{align}
C_1C_2^*+C_1^*C_2=1\,,
\end{align}
which may be used to check the calculation.

To perform the matching, we
note that the derivative of $h_{\mathrm{II}}$ with respect to $z_2$ can be simplified as
\begin{align}
\frac{dh_{\mathrm{II}}}{dz_2}&
=\frac{\sqrt{2\pi}}{\alpha z_2 \mpl}\left(\frac{z_2}{k}\right)^{-\nu}\left[C_1\left(z_2\frac{dJ_{\nu}(z_2)}{dz_2}
-\nu J_{\nu}(z_2)\right)-iC_2\left(z_2\frac{dY_{\nu}(z_2)}{dz_2}
-\nu Y_{\nu}(z_2)\right)\right]\nonumber\\
&
=-\frac{\sqrt{2\pi}}{\alpha\mpl}\left(\frac{z_2}{k}\right)^{-\nu}
\left[C_1J_{\nu+1}(z_2)-iC_2Y_{\nu+1}(z_2)\right]\,,
\end{align}
where in the second step, the relation $xdZ_{\nu}(x)/dx-\nu Z_{\nu}(x)=-xZ_{\nu+1}(x)$
(for $Z=J$ and $Y$) has been used. The matching conditions are 
\begin{align}
h_{\mathrm{I}}(\eta_1)=h_{\mathrm{II}}(\eta_1),\qquad
\frac{dh_{\mathrm{I}}}{dz_1}\bigg|_{\eta=\eta_1}=-\frac{dh_{\mathrm{II}}}{dz_2}\bigg|_{\eta=\eta_1}\,,
\end{align}
from which we obtain the integration constants $C_1$ and $C_2$ as
\begin{align}
C_1&=-\frac{1}{2}\sqrt{\frac{\pi ns}{p}}~e^{ip/s}
\left[\left(\frac{p}{s}+i\right)Y_{\nu+1}\left(\frac{np}{s}\right)
-\frac{ip}{s}Y_{\nu}\left(\frac{np}{s}\right)\right],\label{D1sol}\\
C_2&=\frac{i}{2}\sqrt{\frac{\pi ns}{p}}~e^{ip/s}
\left[\left(\frac{p}{s}+i\right)J_{\nu+1}\left(\frac{np}{s}\right)
-\frac{ip}{s}J_{\nu}\left(\frac{np}{s}\right)\right]\,,\label{D2sol}
\end{align}
where we have introduced the dimensionless wavenumber $p$,
\begin{align}
p\equiv \frac{k}{k_1}\,,
\end{align}
and used the relation $J_{\nu}(x)Y_{\nu+1}(x)-J_{\nu+1}(x)Y_{\nu}(x)=-2/(\pi x)$
to simplify the expressions.
 For the long wavelength modes $k\ll k_1$, the coefficients $C_1$ and $C_2$ can be expanded as
\begin{align}
C_1&=\frac{i~2^{2\nu+1/2}~\Gamma(1+\nu)}{\sqrt{\pi}~(1+2\nu)^{\nu+1/2}}
\left(\frac{s}{p}\right)^{\nu+3/2}e^{ip/s}
\left[1-i~\frac{p}{s}+\frac{(2\nu-3)(2\nu+1)}{16\nu}
\left(\frac{p}{s}\right)^2+\mathcal{O}(p^3)\right],\label{C1app}\\
C_2&=\frac{\sqrt{\pi}~(1+2\nu)^{\nu+3/2}}{2^{2\nu+7/2}~(1+\nu)~\Gamma(1+\nu)}
\left(\frac{p}{s}\right)^{\nu+1/2}~e^{ip/s}\left[\frac{2\nu+3}{2\nu+1}
+i~\frac{p}{s}-\frac{(2\nu+1)(2\nu+7)}{16(\nu+2)}\left(\frac{p}{s}\right)^2
+\mathcal{O}(p^3)\right]\,,\label{C2app}
\end{align}

\subsection{Matching at $\eta=\eta_2$}
Now we consider the matching at the break stage and the second inflationary stage.
 Similar to the case of $h_{\mathrm{II}}$, the solution for the second inflationary 
stage can be expressed as
\begin{align}\label{h3exact}
h_{\mathrm{III}}
&=\frac{\sqrt{2\pi}H_2}{\mpl}\left(\frac{z_3}{k}\right)^{3/2}
\left[-iC_3J_{3/2}(z_3)+C_4Y_{3/2}(z_3)\right]\,,
\end{align}
where $z_3\equiv-k(\eta-\gamma_2)$, and $C_3$ and $C_4$ are the
two constants satisfying the normalization relation,
\begin{align}
 C_3C_4^*+C_3^*C_4=1\,. 
\end{align}
The derivative of $h_{\mathrm{III}}$ with respect to $z_3$ is computed as
\begin{align}
\frac{dh_{\mathrm{III}}}{dz_3}&=\frac{\sqrt{2\pi}H_2}{z_3\mpl }\left(\frac{z_3}{k}\right)^{3/2}\left[-iC_3\left(z_3\frac{J_{3/2}(z_3)}{dz_3}+\frac{3}{2}J_{3/2}(z_3)\right)
+C_4\left(z_3\frac{Y_{3/2}(z_3)}{dz_3}+\frac{3}{2}Y_{3/2}(z_3)\right)\right]\nonumber\\
&=\frac{\sqrt{2\pi}H_2}{\mpl}\left(\frac{z_3}{k}\right)^{3/2}
\left[-iC_3J_{1/2}(z_3)+C_4Y_{1/2}(z_3)\right]\,,
\end{align}
where we have used the relation $xdZ_{\nu}(x)/dx+\nu Z_{\nu}(x)=xZ_{\nu-1}(x)$
at the last step. The matching conditions at $\eta=\eta_2$,
\begin{align}
h_{\mathrm{II}}(\eta_2)=h_{\mathrm{III}}(\eta_2),\qquad
-\frac{dh_{\mathrm{II}}}{dz_2}\bigg|_{\eta=\eta_2}=\frac{dh_{\mathrm{III}}}{dz_3}\bigg|_{\eta=\eta_2}\,,
\end{align}
determine $C_3$ and $C_4$ as
\begin{align}
C_3&=-i\frac{\pi p\sqrt{n}}{2}\bigg\{\big[C_1J_{\nu+1}\left(np\right)-iC_2Y_{\nu+1}\left(np\right)\big]Y_{3/2}(p)-\big[C_1J_{\nu}\left(np\right)-iC_2Y_{\nu}\left(np\right)\big]Y_{1/2}(p)\bigg\},\label{D3sol}\\
C_4&=\frac{\pi p\sqrt{n}}{2}\bigg\{\big[C_1J_{\nu+1}\left(np\right)-iC_2Y_{\nu+1}\left(np\right)\big]J_{3/2}(p)-\big[C_1J_{\nu}\left(np\right)-iC_2Y_{\nu}\left(np\right)\big]J_{1/2}(p)\bigg\}\,.\label{D4sol}
\end{align}
where $C_1$ and $C_2$ are given by \eqref{D1sol} and \eqref{D2sol}, respectively. 

For the long wavelength modes $k\ll k_1$, the coefficients $C_3$ and $C_4$ can be expanded as
\begin{align}
C_3&=\frac{s^{-(\nu+3/2)}}{4\sqrt{2}~(1+\nu)}\left(\frac{s}{p}\right)e^{ip/s}\bigg\{\left(s^{2+2\nu}-1\right)(2\nu+3)-i\big[2\nu+1+s^{2+2\nu}\left(2\nu+3\right)\big]\frac{p}{s}+\xi_1\left(\frac{p}{s}\right)^2+\mathcal{O}(p^3)\bigg\}
,\label{C3app}\\
\nonumber\\
C_4&=-\frac{i~s^{\nu+3/2}}{\sqrt{2}}~e^{ip/s}\left[1-i~\frac{p}{s}+\xi_2\left(\frac{p}{s}\right)^2+\mathcal{O}(p^3)\right]\,,\label{C4app}
\end{align}
where the coefficients $\xi_1$ and $\xi_2$ are given by
\begin{align}
\xi_1&=\frac{1}{4\nu(2+\nu)}
\bigg[~\nu(7+2\nu)(1+2\nu)^2+s^2(2+\nu)(1-2\nu)(3+2\nu)^2\nonumber\\
&\qquad\qquad-s^{2+2\nu}(1+2\nu)(2+\nu)(3+2\nu)(3-2\nu)-s^{4+2\nu}\nu (3+2\nu)(4\nu^2+36\nu+39)\bigg]\,,\label{xi1}\\
\xi_2&=\frac{s^{-\nu+1/2}\pi^{-3/2}}{48\nu(1+2\nu)}
\bigg[~(3+2\nu)^2-s^{2\nu}(1+\nu)(1+2\nu)(3-2\nu)-s^{2+2\nu}\nu (3+2\nu)(5+6\nu)\bigg]\,.\label{xi2}
\end{align}

\subsection{Power spectrum}

The power spectrum of the tensor perturbation may be evaluated by taking the limit
 $z_3\to0$. Inserting the asymptotical behavior of Bessel functions,
\begin{align}
J_{3/2}(z_3\to 0)\to\frac13\sqrt{\frac{2}{\pi}}~z_3^{3/2}\,,
\qquad Y_{3/2}(z_3\to 0)\to-\sqrt{\frac{2}{\pi}}~z_3^{-3/2}\,,
\end{align}
into \eqref{h3exact}, we find
\begin{align}
\lim_{z_3\rightarrow0}h_{\mathrm{III}}=-\frac{2H_2}{\mpl}\frac{C_4}{k^{3/2}}\,,
\end{align}
Hence, the corresponding power spectrum is evaluated as
\begin{align}\label{Ph3exact}
\mathcal{P}_{h}&
=\frac{k^3}{2\pi^2}\sum_{+,\times}\lim_{z_3\rightarrow0}|h_{\mathrm{III}}|^2\nonumber\\
&=\frac{4H_2^2}{\pi^2\mpl^2}\big|C_4\big|^2\nonumber\\
&=\frac{2\mathcal{P}_0}{s^{2(n+1)}}\big|C_4\big|^2\,,
\end{align}
where $\mathcal{P}_0\equiv 2H_1^2/(\pi^2\mpl^2)$,
and we used the definition $s\equiv (H_1/H_2)^{1/(n+1)}$ at the last step.
 The expressions for $C_4$ can be found in \eqref{D4sol}, where $C_1$ and $C_2$ are given in \eqref{D1sol} and \eqref{D2sol}. 
\begin{figure}
\centering
\includegraphics[width=.4\textwidth]{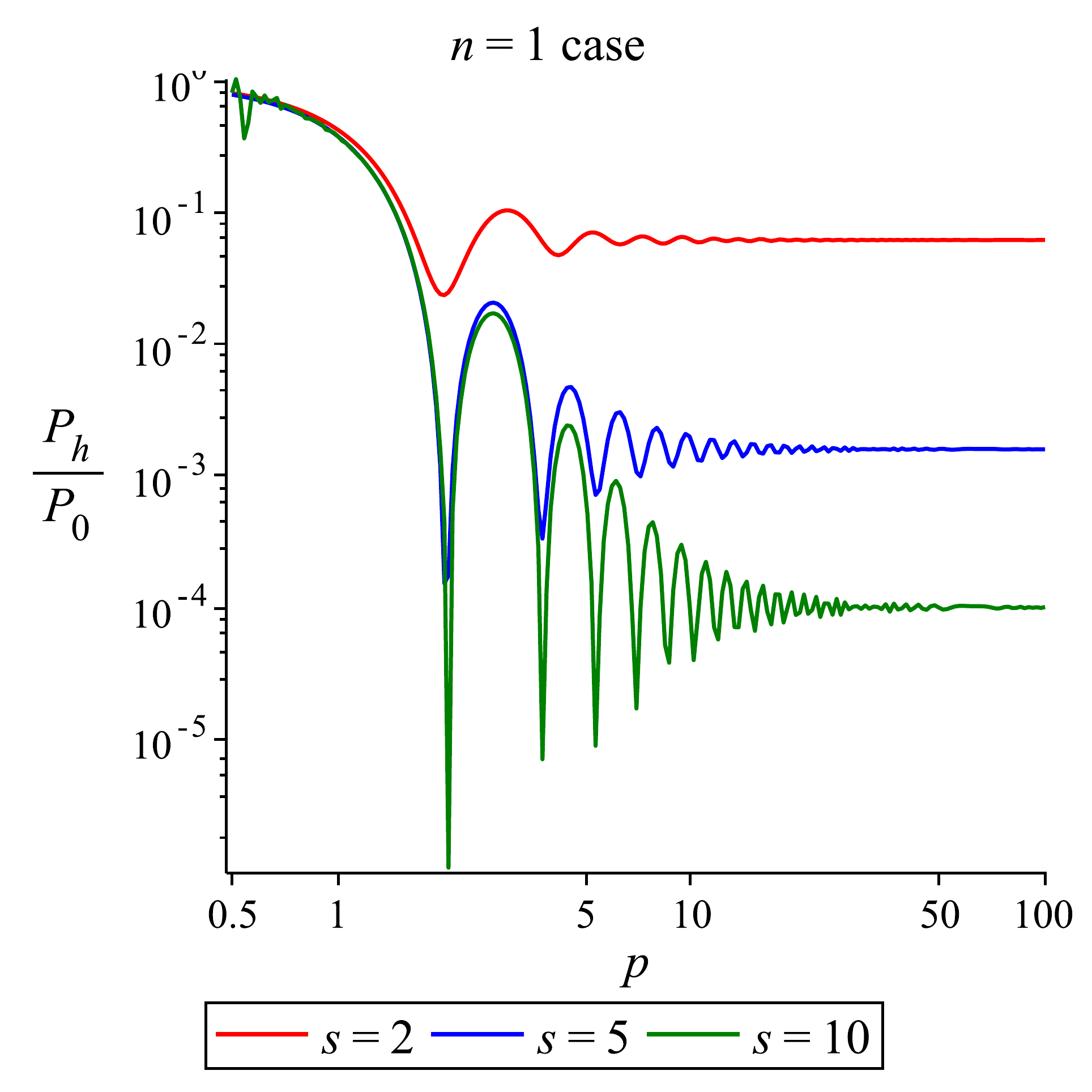}
\includegraphics[width=.4\textwidth]{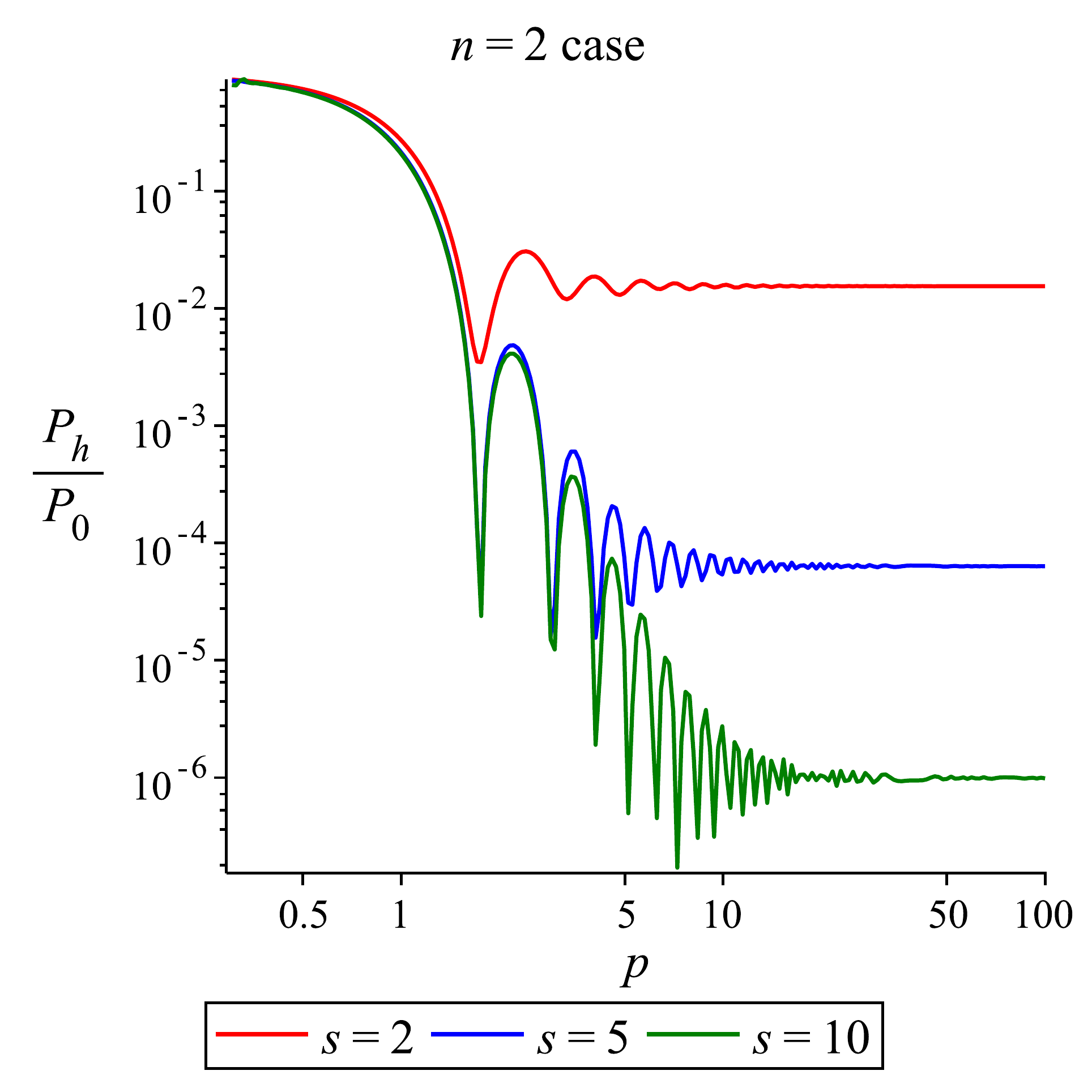}
\caption{The oscillation features in the power spectrum (normalized by $\mathcal{P}_0\equiv 2H_1^2/(\pi^2\mpl^2)$) for the exact solutions for the primordial tensor perturbations \eqref{Ph3exact} with different values of $n$ and $s$. Left: $n=1$ case which corresponds to a radiation-dominated intermediate stage with $w=1/3$. Right: $n=2$ case which corresponds to a matter-dominated intermediate stage with $w=0$. In both cases, as expected, in the limit $p\equiv k/k_1\rightarrow 0$, $\mathcal{P}_{h}\rightarrow\mathcal{P}_0$, while in the limit $p\rightarrow\infty$, $\mathcal{P}_{h}/\mathcal{P}_0\rightarrow (H_2/H_1)^2=s^{-2(n+1)}$. 
However, in contrast to the notion that the perturbation is almost frozen outside the horizon
 in a rapidly expanding universe, the amplitude for the power spectrum decreases 
 as $k$ increases from 0 to $k_1$.}\label{fig:exact}
\end{figure}

In Fig.\ref{fig:exact}, we plot the power spectrum~\eqref{Ph3exact} normalized by 
$\mathcal{P}_0$ for $n=1$ and $n=2$, 
corresponding to a radiation-dominated intermediate stage $w=1/3$ and 
a matter-dominated intermediate stage $w=0$, respectively. 
The modes $k\ll k_1$ present the long wave-length modes which leave the horizon
 at the first stage of inflation and never re-enter the horizon during inflation.
Hence we have $\mathcal{P}_{h}(p\rightarrow0)=\mathcal{P}_0$,
which can be explicitly seen by inserting the 0-th order term of \eqref{C4app} into \eqref{Ph3exact}.
On the other hand, the short wave-length modes $k/k_2=s p\gg1$ never exit the horizon until
 the second inflationary stage. 
 Hence, the power spectrum reduced to  $\mathcal{P}_{h}(p\gg 1)
 =\mathcal{P}_0(H_2/H_1)^2=\mathcal{P}_0s^{-2(n+1)}$. 
There appears an oscillatory behavior for modes $k_1<k<k_2$.
The result indicates that the characteristic frequency depends only on the
EOS of the break stage, but independent of $s$ for large $s$. 
A larger $s$ implies a longer intermediate period, and hence a larger value of $k_2-k_1$,
thus resulting in a wider spread of the oscillation behavior with a fixed frequency.

A rather unexpected feature in the spectrum we find on the very large scales $k<k_1$ is the
existence of a fairly strong suppression relative to $\mathcal{P}_0$, as seen in Fig.\ref{fig:exact}.
Since these modes never re-enter the horizon during the entire stage, one would expect
them to stay almost constant in time after they left the horizon at the first stage.
However, we find that the amplitude for the power-spectrum decreases appreciably 
as $p$ increases from 0 to 1. For example, for $n=s=2$ case, we have $\mathcal{P}_{h}^{\mathrm{exact}}(p=1)\approx0.278\mathcal{P}_0$. 

Apparently this feature is due to the non-trivial evolution of the modes on superhorizon scales.
 In fact, using the expression for $C_4$ expanded up through $O(p^2)$,  \eqref{C4app},
 the power spectrum \eqref{Ph3exact} for the long wavelength modes $k<k_1$ is
 evaluated as  
\begin{align}\label{Ph3exactk2}
\frac{\mathcal{P}_{h}}{\mathcal{P}_0}&
=1-\frac{2\nu+3}{24\nu(1+\nu)}\left[s^{2}\nu(5+6\nu)-3(2\nu-1)(\nu+1)
-s^{-2\nu}(2\nu+3)\right]\left(\frac{p}{s}\right)^2+\mathcal{O}\left(p^4\right)\,.
\end{align}
Later we clarify the source of this $O(p^2)$ correction by appealing to an approximate method
in which all the functions involved are elementary functions. At the same time we find
that the amplitude at $p=1$ is surprisingly well approximated by taking into account only the
$O(p^2)$ correction term and taking the limit $p\to1$ of it.
We also show that the $O(p^2)$ correction term is negative for any EOS with $w>-1/3$ (see Appendix~\ref{nogo}). 
Hence the spectrum is always suppressed relative to ${\mathcal{P}_0}$ irrespective of 
the EOS as long as the intermediate stage is a decelerating universe.

\section{Approximate method}\label{WKBsec}
What we have obtained so far are based on the exact solutions of the tensor perturbation.
However, since they involve special functions, namely, the Bessel functions, it is not
necessarily easy to understand the physical picture behind the result.
In this section, we employ an approximate method for solving the mode functions
in order to clarify the essence of the result. At the same time, by comparing the
result obtained by the approximate method with the exact result, we justify its validity.

First we prepare the initial mode function. The EOM for the tensor mode function
is given by Eq.~(\ref{eomh}) with the normalization condition~(\ref{KGnorm}),
which we recapitulate:
\be\label{basicEOM}
h''+2\mathcal Hh'+k^2h=0\,; \quad
\left(\frac{\mpl}{2}\right)^2\left(h\bar{h}'-\bar{h}h'\right)=\frac{i}{a^2}\,.
\ee
When the mode is well inside the horizon $k\gg\mathcal H$,
the equation can be solved by the WKB approximation,
\be\label{h<}
h\xrightarrow{~k\gg{\cal H}}\frac{1}{a(\eta)}\left(c_1e^{-ik\eta}+c_2e^{ik\eta}\right)\,,
\ee
where $c_1$ and $c_2$ are constants. Imposing the condition that
the initial positive frequency mode function should behave as the Minkowski one,
we immediately find $c_2=0$, and obtain
\be\label{h<BD}
h\xrightarrow{~\eta\rightarrow-\infty~}\frac{2}{\mpl}\frac{e^{-ik\eta}}{\sqrt{2k}a(\eta)}.
\ee
Apart from an irrelevant phase, this coincides with Eq.~(\ref{h1sol}) in the limit 
$z_1=-k(\eta-\gamma)\gg1$.

When the wavelength exceeds the Hubble horizon size, the solution is
approximately given by
\be\label{h>}
h\xrightarrow{~k\ll {\cal H}~} d_1+d_2\int^\eta\frac{d\eta}{a^{2}(\eta)}.
\ee
where $d_1$ and $d_2$ are constants. In the conventional single-stage inflation,
it is commonly the case that the contribution of the second decaying solution is
negligible. In such a case one can put $d_2=0$ and simply join the WKB solution
with the constant solution $d_1$ at the horizon crossing. Thus setting $a=k/H$
in the WKB mode function~(\ref{h<BD}), one obtains the standard result,
\begin{align}
h\xrightarrow{~k\ll {\cal H}~} d_1=\frac{2H}{\mpl\sqrt{2k^3}}\,,
\end{align}
where we have put the irrelevant phase to zero for simplicity.

However, as we have seen in the previous section, we found that the effect of the
non-trivial evolution outside the horizon plays an important role in the determining
the amplitude of the power spectrum. Hence, in the analysis below, we will take into
 account both the correction term of order $(k/{\cal H})^2$ to the constant solution
 and the decaying solution at the leading order. As we will see later,
 this rather simple-minded approximation turns out to be unexpectedly accurate.

\subsection{Intermediate wavelength: $k_1<k<k_2$}
\begin{figure}
\centering
\includegraphics[width=.7\textwidth]{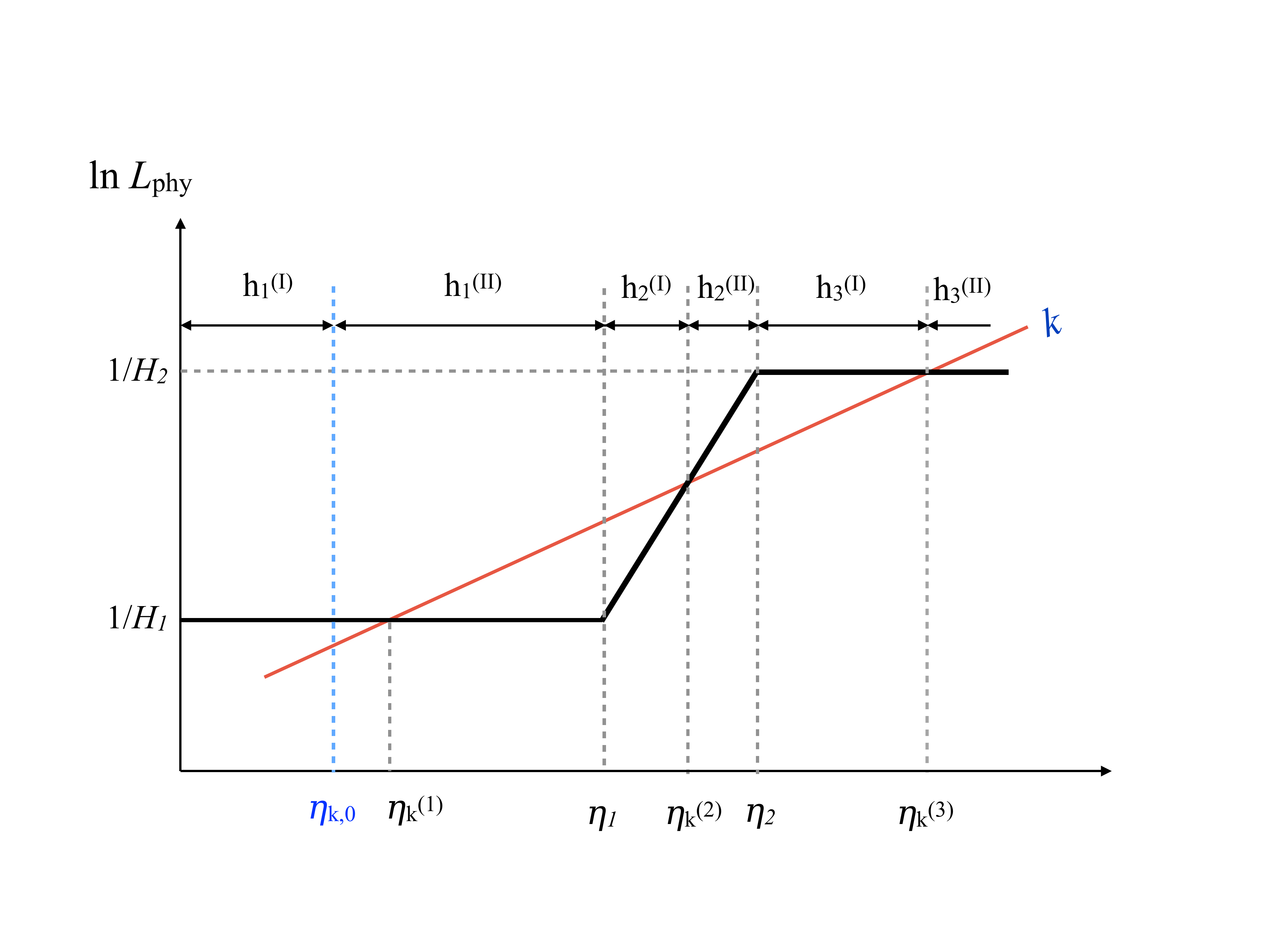}
\caption{Schematic diagram to demonstrate both the boundaries for matching conditions and the corresponding different pieces of solutions for the modes with intermedia wavenumber $k_2>k>k_1$. As explained below \eqref{h1out}, in order to match with the long wavelength modes so that the asymptotic behavior of the power spectrum $P(k/k_1\rightarrow0)/P_0=1$, we choose the matching point for $h_1^{(\mathrm{I})}$ and $h_1^{(\mathrm{II})}$ located at $\eta_{k,0}$, which is slightly earlier than $\eta_k^{(1)}$.}\label{fig:WKBinter}
\end{figure}

We first discuss the most complicated case of the intermediate wavenumbers
$k_1<k<k_2$. As shown in Fig.~\ref{fig:WKBinter}, we have to consider the matching 
at five points in this case: 
$\eta_k^{(1)}$, $\eta_1$, $\eta_k^{(2)}$, $\eta_2$ and $\eta_k^{(3)}$,
where $\eta_k^{(1)}$, $\eta_k^{(2)}$, and $\eta_k^{(3)}$ are the epochs of 
the first horizon exit, horizon re-entry, and the second horizon exist, respectively.

\subsubsection{Matching at $\eta_k^{(1)}$}
At the stage  $\eta<\eta_k^{(1)}$, the WKB solution is
\be\label{h1wkb}
h_1^{(\mathrm{I})}(\eta)\approx\frac2\mpl\frac{e^{-ik(\eta-\gamma_1)}}{\sqrt{2k}~a_{\mathrm{I}}}= We^{iz_1}z_1\,;\quad z_1\equiv-k(\eta-\gamma_1)\,,
\ee
where
\be
W\equiv\frac{\sqrt{2}H_1}{k^{3/2}\mpl}\,.
\ee
At $\eta>\eta_k^{(1)}$, we employ the long wavelength expansion and take into account
the  $O(z_1^2)$ correction to the constant solution, and the decaying mode proportional to 
$z_1^3$. As given in Appendix \ref{WKBsuppress}, it is expressed as 
\be\label{h1out}
h_1^{(\mathrm{II})}(\eta)=W\left[A\left(1+\frac{z_1^2}{2}\right)+Bz_1^3\right]\,,
\ee
where $A$ and $B$ are constants.
Note that $W$ is the amplitude one would obtain on superhorizon scales
in a single-stage inflation. Thus one should get $|A|=1$ if there were no intermediate
break stage.

Naively one would think that the matching is to be done at the exact horizon crossing
point $\eta=\eta_k^{(1)}$. 
 However, since we have introduced the correction terms in \eqref{h1out}, 
 the tensor perturbations are not exactly conserved outside the horizon.
This would then give rise to a correction to the amplitude $A$ even in the limit $k/k_1\to0$,
and lead to an incorrect result.
To remedy this problem, we adopt the following prescription. Namely,
we match the WKB solution with the long wavelength solution at $\eta=\eta_{k,0}$
slightly different from $\eta_k^{(1)}$ so that we have $|A|=1$.
Specifically, solving the matching conditions
 $h_1^{(\mathrm{I})}=h_1^{(\mathrm{II})}$ and $dh_1^{(\mathrm{I})}/d\eta=dh_1^{(\mathrm{II})}/d\eta$
 at $\eta=\eta_{k,0}$, the constants $A$ and $B$ are obtained as
\begin{align}\label{ABexp}
A=\frac{2z_0\left(2-iz_0\right)}{z_0^2+6},\qquad B=\frac{2-z_0^2+iz_0\left(z_0^2+2\right)}{z_0^2\left(z_0^2+6\right)}\,,
\end{align}
where $z_0\equiv z_1(\eta_{k,0})$.  Then requiring $|A|=1$, we find
\begin{align}\label{z0exp}
 z_0=\left[\frac{2}{3}\left(2\sqrt{7}-1\right)\right]^{1/2}\approx1.69\,.
\end{align}
As clear from this, the matching is done slightly inside the horizon. 
With the above `renormalization' of the amplitude $A$, we obtain
the approximate solution outside the horizon \eqref{h1out} with
the coefficients given by \eqref{ABexp} and \eqref{z0exp}.

\subsubsection{Matching at $\eta_1$}

Now let us discuss the matching of $h_1^{(\mathrm{II})}$ and $h_2^{(\mathrm{I})}$. 
In the range $\eta_1<\eta<\eta_k^{(2)}$, as shown in Appendix \ref{WKBsuppress}, 
taking into account the correction term and decaying mode, we have
\begin{align}\label{h2WKB1}
h_2^{(\mathrm{I})}=W\left[C\left(1-\frac{z_2^2}{2(2n+1)}\right)+Dz_2^{1-2n}\right]\,.
\end{align}
The decaying mode proportional to $z_2^{1-2n}$ is introduced to satisfy the matching conditions.
From the matching conditions $h_1^{(\mathrm{II})}|_{\eta_1}=h_2^{(\mathrm{I})}|_{\eta_1}$ 
and $dh_1^{(\mathrm{II})}/d\eta|_{\eta_1}=dh_2^{(\mathrm{I})}/d\eta|_{\eta_1}$, 
the constants $C$ and $D$ are obtained as
\begin{align}
C&=\frac{As\left[2s^2(2n-1)-p^2\right]-2B(n+1)p^3}{s\left[2s^2(2n-1)-n^2p^2\right]}
\label{C}\,,\\
D&=p\left(\frac{np}{s}\right)^{2n}\frac{As(n+1)(np^2-2s^2)
	+Bp\left[n(3n+2)p^2-6(2n+1)s^2\right]}{s^2(2n+1)\left[n^2p^2+2s^2(1-2n)\right]}\,,
\label{D}
\end{align}
where $A$ and $B$ are given by \eqref{ABexp} with \eqref{z0exp}. 

\subsubsection{Matching at $\eta_k^{(2)}$}

The mode re-enters the horizon at $\eta=\eta_k^{(2)}$. After that epoch,
we approximate it by the WKB mode function, 
\be\label{WKB1}
h_2^{(\mathrm{II})}(\eta)
=\frac{W}{\left[k(\eta-\beta)\right]^n}\left(c_1e^{-ik(\eta-\beta)}+c_2e^{ik(\eta-\beta)}\right)\,;
 \qquad~\eta_k^{(2)}<\eta<\eta_2.
\ee
We match this solution and its derivative with the solution \eqref{h2WKB1} at $\widetilde{\eta}_k^{(2)}$, defined as 
\begin{align}
z_2(\widetilde{\eta}_k^{(2)})\equiv\tilde{z}_0z_2(\eta_k^{(2)})= \tilde{z}_0n,
\end{align}
where $\tilde{z}_0$ is a parameter of $O(1)$, which is inserted to keep track of the effect 
of the choice of the matching point, but which will be set to $\tilde{z}_0=1$ in the end 
when we numerically compare the approximate result with the exact one. 
The matching conditions give
\begin{align}
h_2^{(\mathrm{I})}(\widetilde{\eta}_k^{(2)})
=h_2^{(\mathrm{I})}(\widetilde{\eta}_k^{(2)})\qquad&\Longrightarrow\qquad \widetilde{c}_1+\widetilde{c}_2=(n \tilde{z}_0)^n\left[C\left(1-\frac{n^2\tilde{z}_0^2}{2(1+2n)}\right)+D(n\tilde{z}_0)^{1-2n}\right],\label{e1}\\
h_2^{(\mathrm{II})\prime}(\widetilde{\eta}_k^{(2)})
=h_2^{(\mathrm{I})\prime}(\widetilde{\eta}_k^{(2)})\qquad&\Longrightarrow\qquad \widetilde{c}_1\left(\frac{1}{\tilde{z}_0}+i\right)+\widetilde{c}_2\left(\frac{1}{\tilde{z}_0}-i\right)=(n\tilde{z}_0)^n\left[\frac{Cn\tilde{z}_0}{1+2n}+(2n-1)D(n\tilde{z}_0)^{-2n}\right].\label{e2}
\end{align}
where for notational simplicity, we have introduced
\begin{align}
\widetilde{c}_1&\equiv c_1e^{-in\tilde{z}_0},\qquad \widetilde{c}_2\equiv c_2e^{in\tilde{z}_0}.
\end{align}
It is straightforward to solve \eqref{e1} and \eqref{e2} for $\widetilde{c}_1$ and $\widetilde{c}_2$,
\begin{align}
\widetilde{c}_1&=\frac{(n\tilde{z}_0)^n}{2}\bigg\{D\tilde{z}_0(n\tilde{z}_0)^{1-2n}-\frac{C}{2(1+2n)}\tilde{z}_0\left(n^2\tilde{z}_0-4n-2\right)\nonumber\\
&\qquad\qquad\qquad+i\left[D\tilde{z}_0(n\tilde{z}_0)^{-2n}(1-n)+C\tilde{z}_0\left(1-\frac{n(2+n)\tilde{z}_0^2}{2(1+2n)}\right)\right]\bigg\},\label{c1}\\
\widetilde{c}_2&=\widetilde{c}_1^{~*},\label{c2}
\end{align}
where $C$ and $D$ are given in \eqref{C} and \eqref{D}.

\subsubsection{Matching at $\eta_2$}

Now consider the matching at $\eta=\eta_2$. Actually, mathematically speaking
it is unnecessary to do this matching because the WKB solution in the form \eqref{h<} is 
valid irrespective of the background expansion law. So the only thing we have to do
is to match the scale factor and the phase of the mode function.

To do so, we rewrite the WKB solution \eqref{WKB1} at $\eta<\eta_2$ in the form
 \eqref{h<} by noting the expression of the scale factor at the break stage given in \eqref{a-eta}.
 This gives
 \begin{align}
 h_2^{(\mathrm{II})}(\eta)
 =\frac{\alpha}{k^n}\frac{W}{a_{\rm{II}}(\eta)}\left(c_1e^{-ik(\eta-\beta)}+c_2e^{ik(\eta-\beta)}\right)\,.
 \end{align}
With this form, it is straightforwardly matched to the WKB solution at $\eta>\eta_2$
by simply replacing the scale factor $a_{\rm{II}}$ by that at the second inflationary stage,
$a=a_{\rm{III}}$,
\begin{align}
h_3^{(\mathrm{I})}(\eta)
&=\frac{\alpha}{k^n}\frac{W}{a_{\rm{III}}(\eta)}\left(c_1e^{-ik(\eta-\beta)}+c_2e^{ik(\eta-\beta)}\right)
\nonumber\\
&=WVk(\eta-\gamma_2) \left(c_3e^{-ik(\eta-\gamma_2)}+c_4e^{ik(\eta-\gamma_2)}\right)\,;
\quad V=-\frac{1}{n^np^{n+1}}\,,
\label{h3fin}
\end{align}
where 
\begin{align}
c_3&=c_1e^{-ik(\gamma_2-\beta)}=c_1e^{-ip(n+1)}=\widetilde{c}_1e^{-ip(n+1)+in\tilde{z}_0}\,,
\nonumber\\
c_4&=c_2e^{ik(\gamma_2-\beta)}=c_2e^{ip(n+1)}=\widetilde{c}_2e^{-ip(n+1)+in\tilde{z}_0}\,.
\end{align}

\subsubsection{Matching at $\eta_k^{(3)}$}
 
For this matching, similar to the superhorizon solution at the first inflationary stage ,
instead of assuming a constant solution, we use a slightly more accurate 
solution at $\eta>\eta_k^{(3)}$,
\be\label{h3out}
h_3^{(\mathrm{II})}(\eta)=W\left[F\left(1+\frac{z_3^2}{2}\right)+Gz_3^3\right]\,.
\ee
The solution \eqref{h3out} should be matched with \eqref{h3fin}  at horizon exit
 $\eta=\widetilde{\eta}_k^{(3)}$, defined by
\begin{align}
z_3(\widetilde{\eta}_k^{(3)})\equiv y_0z_3(\eta_k^{(3)})=y_0,
\end{align}
 where $y_0$ is a parameter
 of $O(1)$. Similarly as the case for $\tilde{z}_0$, we will set $y_0=1$ at the end for numerical comparison of approximate result with exact one. The matching conditions are
 $h_3^{(\mathrm{II})}(\widetilde{\eta}_k^{(3)}) =h_3^{(\mathrm{I})}(\widetilde{\eta}_k^{(3)})$
 and $h_3^{(\mathrm{II})\prime}(\widetilde{\eta}_k^{(3)})
 =h_3^{(\mathrm{I})\prime}(\widetilde{\eta}_k^{(3)})$, which respectively give 
\begin{align}
F\left(1+\frac{y_0^2}{2}\right)+Gy_0^3
&=-Vy_0\left(\widetilde{c}_1e^{-i\psi}+\widetilde{c}_2e^{i\psi}\right),
\label{eqn:F1}\\
Fy_0+3Gy_0^2
&=-V\left[\widetilde{c}_1e^{-i\psi}(1+iy_0)+\widetilde{c}_2e^{i\psi}(1-iy_0)\right],
\label{eqn:F2}
\end{align}
where $\psi\equiv(1+n)p-n\tilde{z}_0-y_0$. Hence we obtain
\begin{align}\label{Fexp}
F=-\frac{2Vy_0}{6+y_0^2}\left[(2-iy_0)\widetilde{c}_1e^{-i\psi}
+(2+iy_0)\widetilde{c}_2e^{i\psi}\right]\,,
\end{align}
where $\widetilde{c}_1$ and $\widetilde{c}_2$ are given by \eqref{c1} and \eqref{c2}, respectively.

 Finally, the power spectrum at the end of inflation is obtained as
\begin{align}\label{Ph3int}
\mathcal{P}_{k_1<k<k_2}
=\lim_{z_3\rightarrow0}\left(\frac{k^3}{2\pi^2}\sum_{+,\times}|h_3^{(\mathrm{II})}|^2\right)
=\mathcal{P}_{0}\left|F\right|^2\,,
\end{align}
where $\mathcal{P}_0\equiv 2H_1^2/(\pi^2\mpl^2)$.

\subsection{Long wavelength: $k<k_1$}
For the modes $k<k_1$, the difference from the previous case is that
they never re-enter the horizon. Thus the superhorizon solution at the intermediate stage
$h_2^{(\mathrm{I})}$ is directly matched to the superhorizon solution at the second inflationary
stage $h_3^{(\mathrm{II})}$.
Similar to \eqref{h3out}, we express $h_3^{(\mathrm{II})}$ as
\be\label{h3long}
h_{3, k<k_1}^{(\mathrm{II})}(\eta)
=W\left[\widetilde{F}\left(1+\frac{z_3^2}{2}\right)+\widetilde{G}z_3^3\right]\,.
\ee
Matching this solution with $h_2^{(\mathrm{I})}$ given by \eqref{h2WKB1}
at $\eta_2$, we have
\begin{align}
\widetilde{F}\left(1+\frac{p^2}{2}\right)+\widetilde{G}p^3
&=C\left[1-\frac{n^2p^2}{2(1+2n)}\right]+D(np)^{1-2n},
\label{eqn:F1}\\
 \widetilde{F}p+3\widetilde{G}p^2&=\frac{C np}{1+2n}+D(2n-1)(np)^{-2n},
 \label{eqn:F2}
\end{align}
which are solved for $\widetilde{F}$ and $\widetilde{G}$ to give
\begin{align}
\widetilde{F}&
=\frac{2}{p^2+6}\left[C\left(3-\frac{(2+3n)np^2}{2(1+2n)}\right)+D(n+1)n^{-2n}p^{1-2n}\right],
\label{Flongexp}\\
\widetilde{G}&
=-\frac{1}{p^2\left(p^2+6\right)}\left[C\left(2p-\left(\frac{n}{1+2n}\right)\left(1+(1+n)p^2\right)\right)
+D(p^2-4+2)(np)^{-2n}\right].
\label{Glongexp}
\end{align}
The power spectrum at $k<k_1$ is given by
\begin{align}\label{Ph3long}
\mathcal{P}_{k<k_1}&=\lim_{z_3\rightarrow0}\left(\frac{k^3}{2\pi^2}\sum_{+,\times}|h_{3, k<k_1}^{(\mathrm{II})}|^2\right)=\mathcal{P}_{0}\left|\widetilde{F}\right|^2\,.
\end{align}
In the limit $p= k/k_1\rightarrow0$, from \eqref{C} and \eqref{D} , we have $C=A$ and 
$D\propto Ap^{2n+1}$. Inserting these into \eqref{Flongexp} and \eqref{Glongexp}, 
using \eqref{h3long}, 
we obtain the asymptotic behavior,
\begin{align}
\lim_{k/k_1\rightarrow0}\mathcal{P}_{k<k_1}=\mathcal{P}_0\left|A\right|^2=\mathcal{P}_0\,.
\end{align}
This justifies the choice $\eta=\eta_{k,0}$ set by the condition $|A|=1$
as the matching point instead of $\eta=\eta_k^{(1)}$, as argued below \eqref{h1out}.

We note that for these long wavelength modes, 
$\widetilde{F}$ can be expanded to $k^2$ as
\begin{align}\label{Flongser}
\widetilde{F}&=
A\left\{1-\frac{2\nu+3}{48\nu(1+\nu)}
\left[s^{2}\nu(5+6\nu)-3(2\nu-1)(\nu+1)-s^{-2\nu}(2\nu+3)\right]
\left(\frac{p}{s}\right)^2+\mathcal{O}\left(p^4\right)\right\}\,.
\end{align}
Hence we obtain
\begin{align}\label{Ph3longWKBser}
\frac{\mathcal{P}_{k<k_1}}{\mathcal{P}_{0}}=|\widetilde{F}|^2=
1-\frac{2\nu+3}{24\nu(1+\nu)}\left[s^{2}\nu(5+6\nu)-3(2\nu-1)(\nu+1)
-s^{-2\nu}(2\nu+3)\right]\left(\frac{p}{s}\right)^2+\mathcal{O}\left(p^4\right)\,,
\end{align}
which exactly recovers the power spectrum to $O(k^2)$, \eqref{Ph3exactk2},
obtained from the exact solution. This proves our argument that the suppression of 
the power spectrum at $k\leq k_1$ comes from the correction 
terms proportional to $k^2$. 

Furthermore, provided that the background of the break stage (between the two inflationary stages)
describes an expanding universe, it can be proved that the correction term proportional 
to $p^2$ in \eqref{Ph3longWKBser} is always negative (see a proof in Appendix~\ref{nogo}).
Thus, a suppression of the tensor perturbation power spectrum
at those long wavelengths that left horizon during the first stage of inflation
is a generic feature of inflationary models with an intermediate break.

\subsection{Short Wavelength: $k>k_2$}
Now we consider the modes that keep staying inside horizon until the second stage of inflation.
In this case, the WKB solution \eqref{h<BD} is valid until the horizon when $\eta=\eta_k=-1/k$.
Here we recapitulate it:
  \begin{align}
  h_{\rm{WKB}}(\eta)=\frac{2}{\mpl}\frac{e^{-ik\eta}}{\sqrt{2k}a(\eta)}\,.
\label{WKBsol}
  \end{align}
Therefore we may use the conventional, leading order approximation to match the
amplitude of the WKB solution with a constant solution outside the horizon.
Namely, 
\begin{align}
h(\eta>\eta_k)=h_{\rm{WKB}}(\eta_k)= \frac{2}{\mpl}\frac{H_2}{\sqrt{2k^3}}\,,
\end{align}
apart from an irrelevant phase.
We see that the intermediate stage does not affect the 
short wavelength modes at all, as which never exit horizon until the second inflationary stage, 
which coincides our physical understanding.

The power spectrum is given by
\be\label{temp6}
\mathcal P_h=\frac{2H_2^2}{\pi^2\mpl^2}=\frac{2H_1^2}{\pi^2s^{2(n+1)}\mpl^2}\,,
\ee
where $s\equiv(H_1/H_2)^{n+1}$ has been used in the final step.

\begin{figure}
\centering
\includegraphics[width=.38\textwidth]{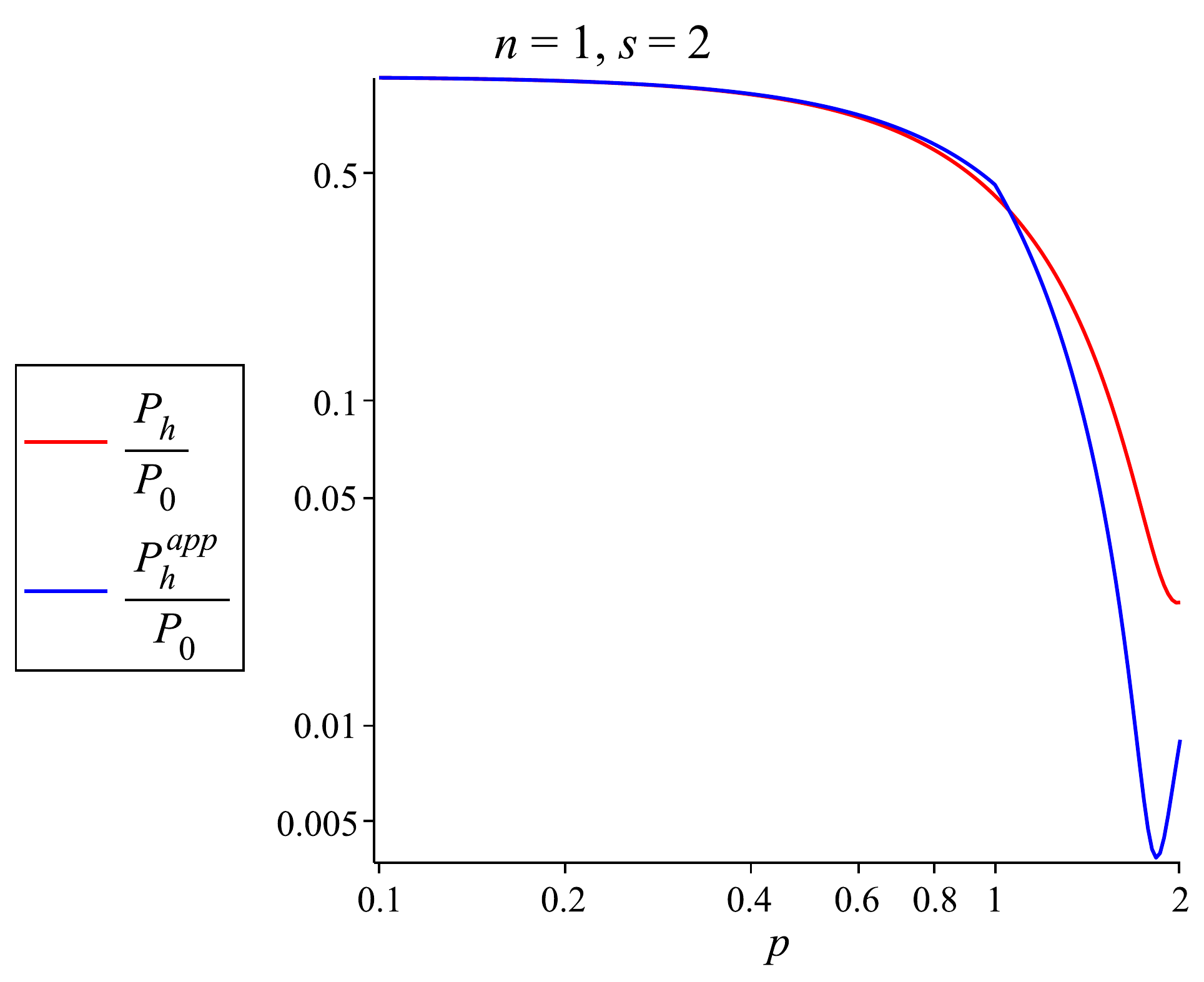} 
\includegraphics[width=.3\textwidth]{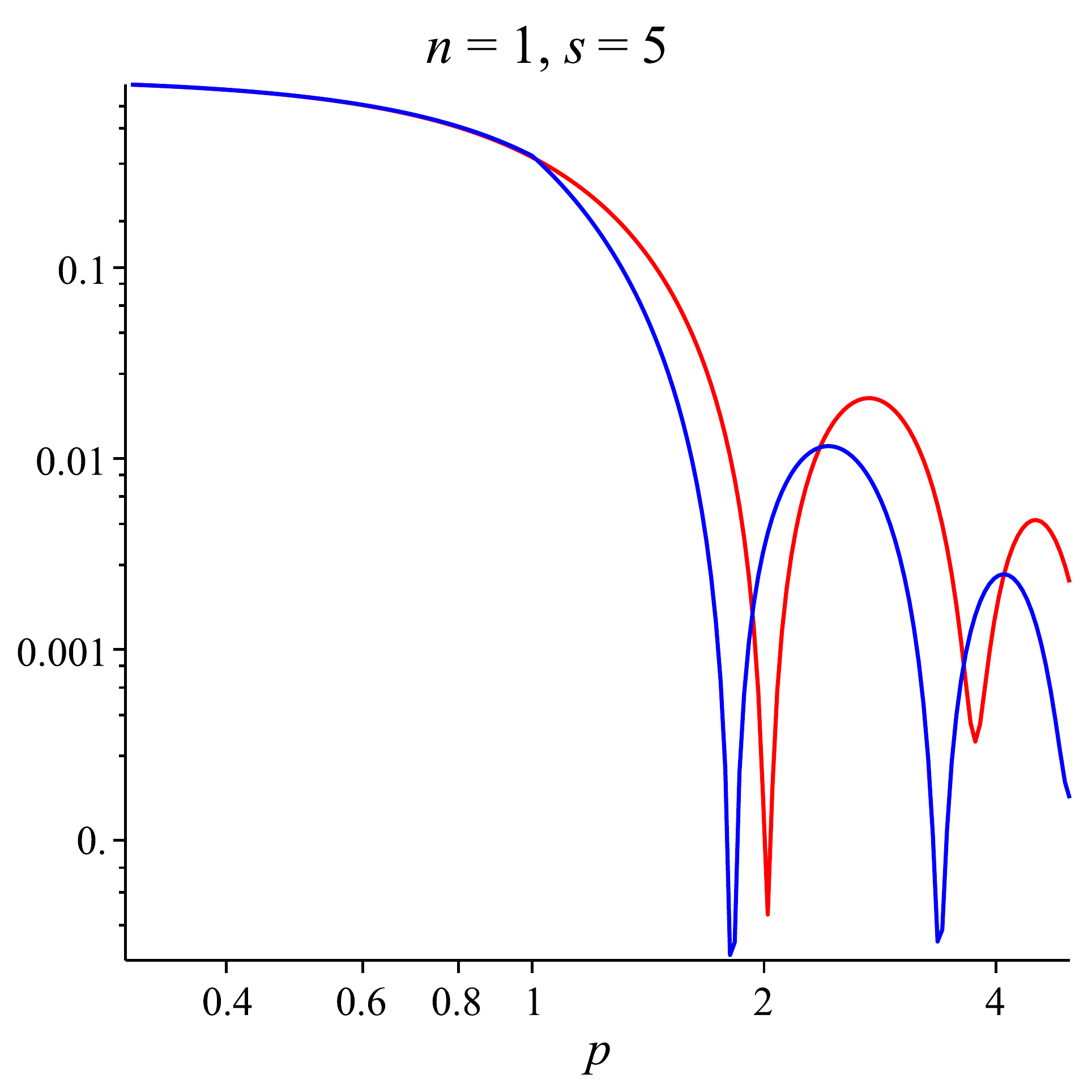}
\includegraphics[width=.3\textwidth]{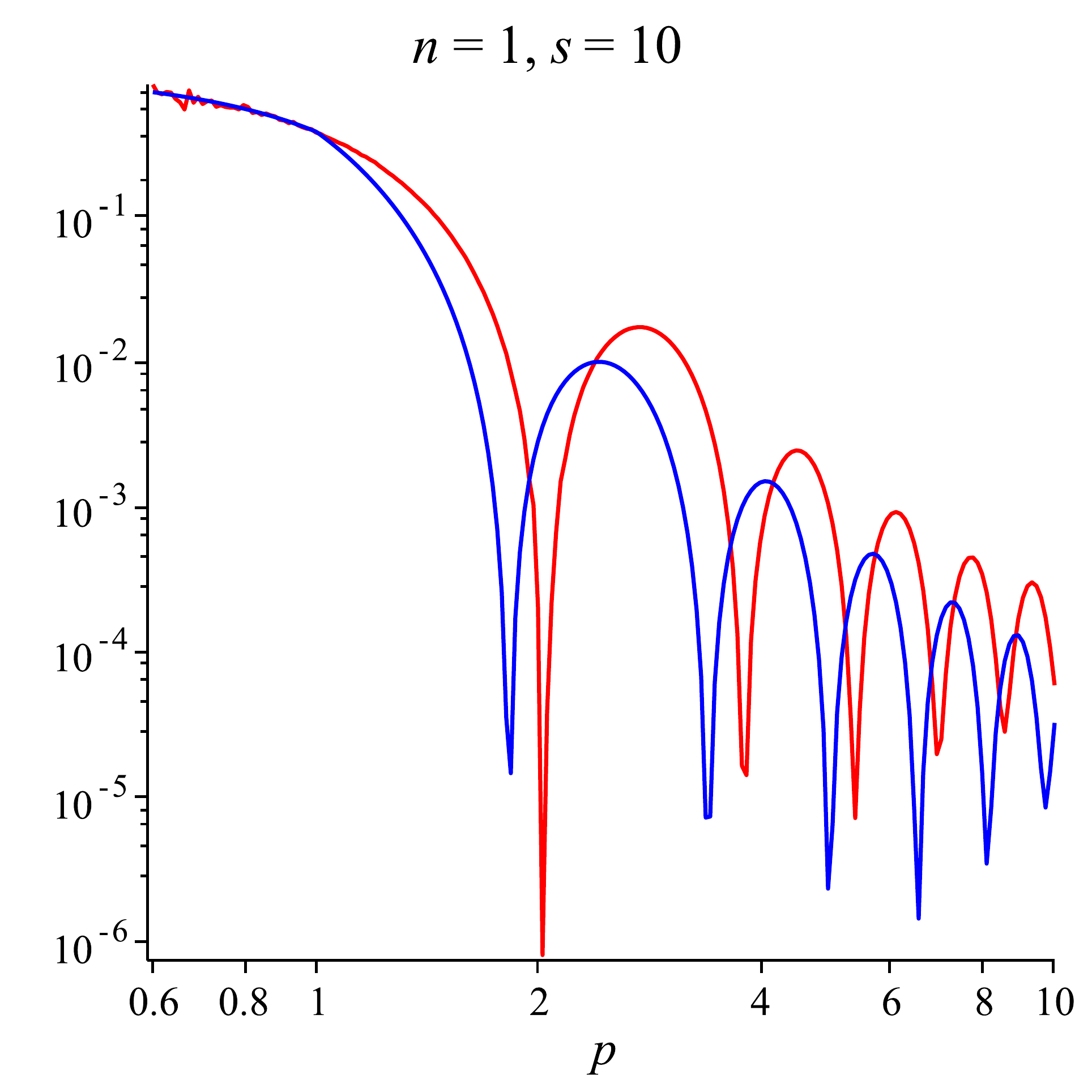}
\includegraphics[width=.38\textwidth]{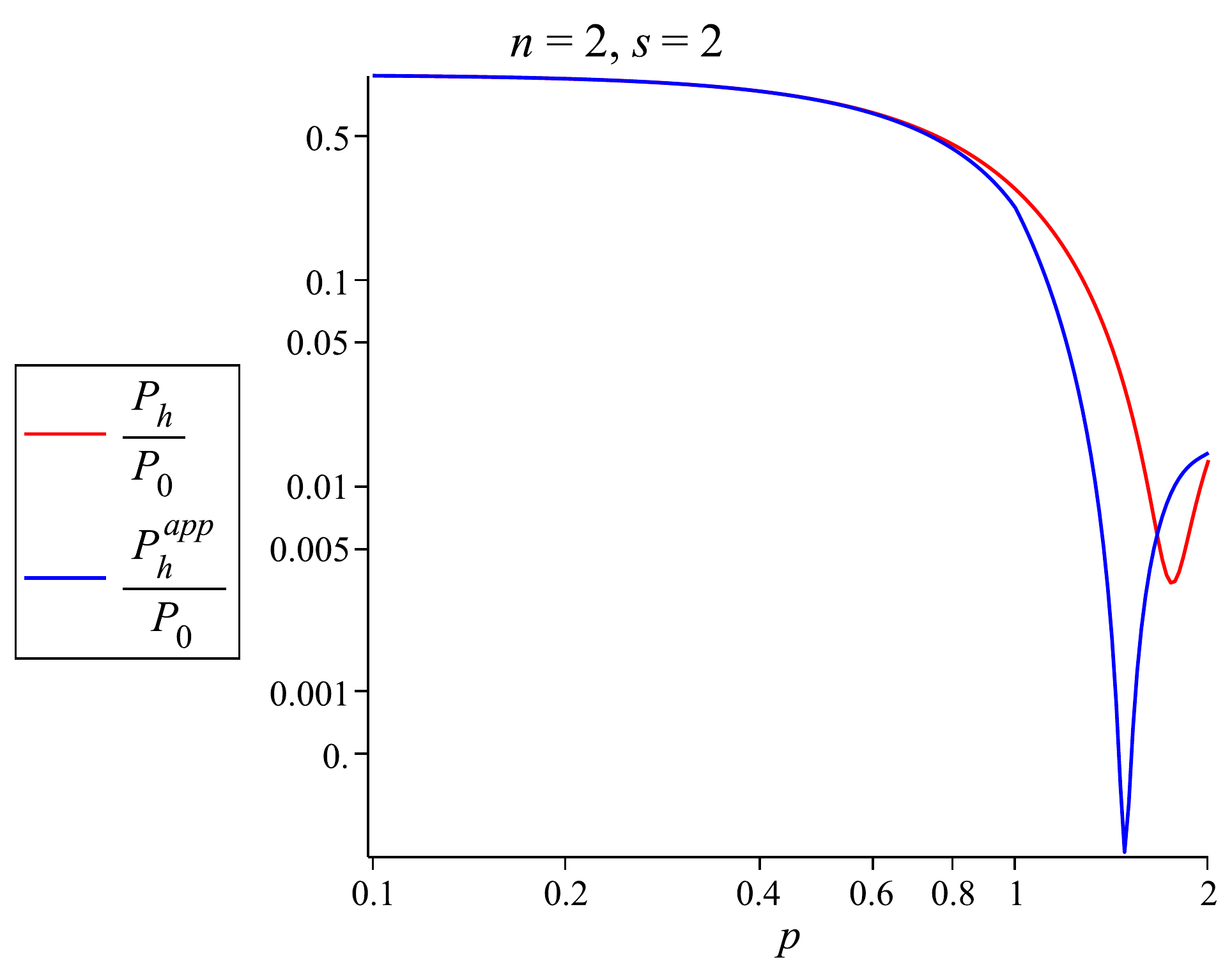}
\includegraphics[width=.3\textwidth]{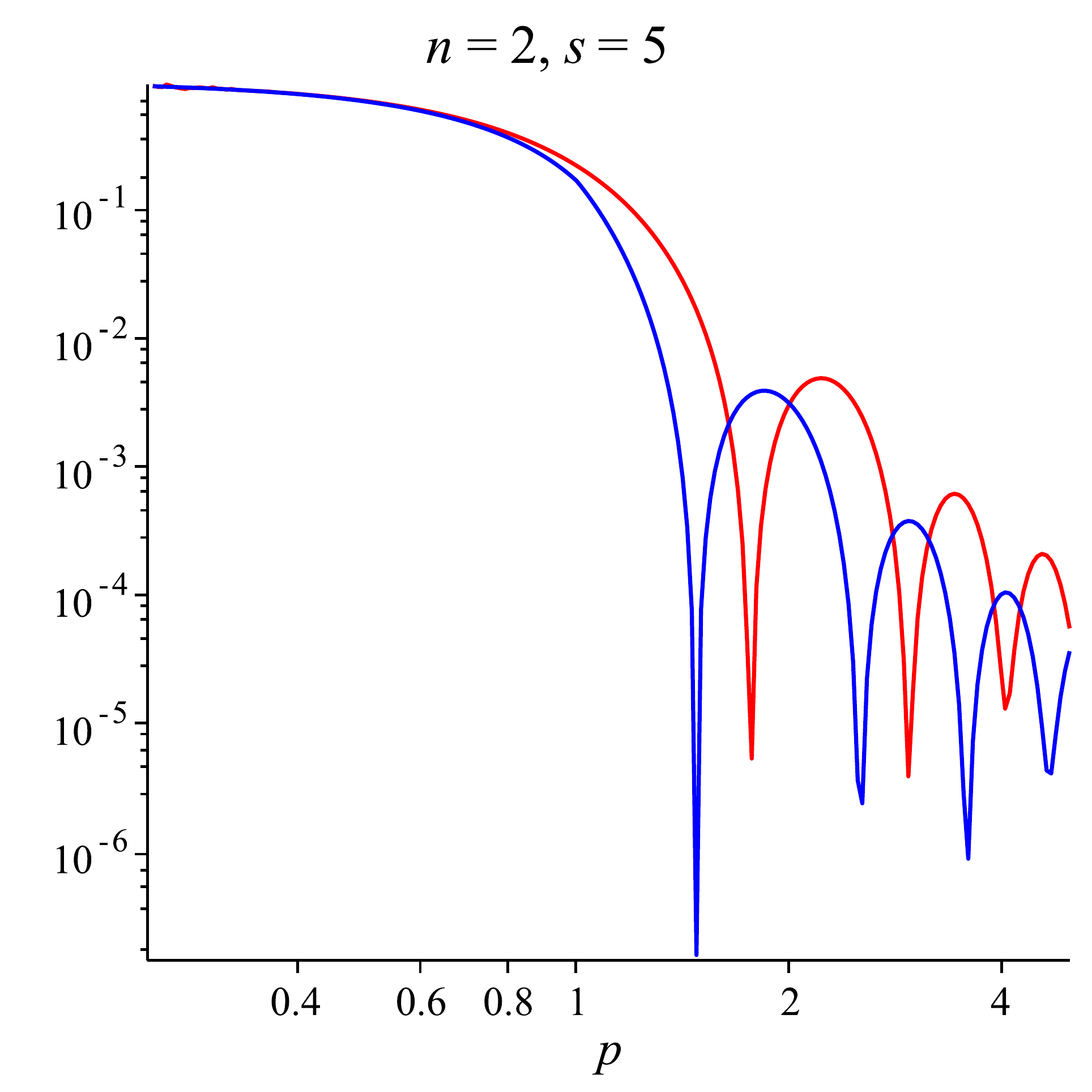}
\includegraphics[width=.3\textwidth]{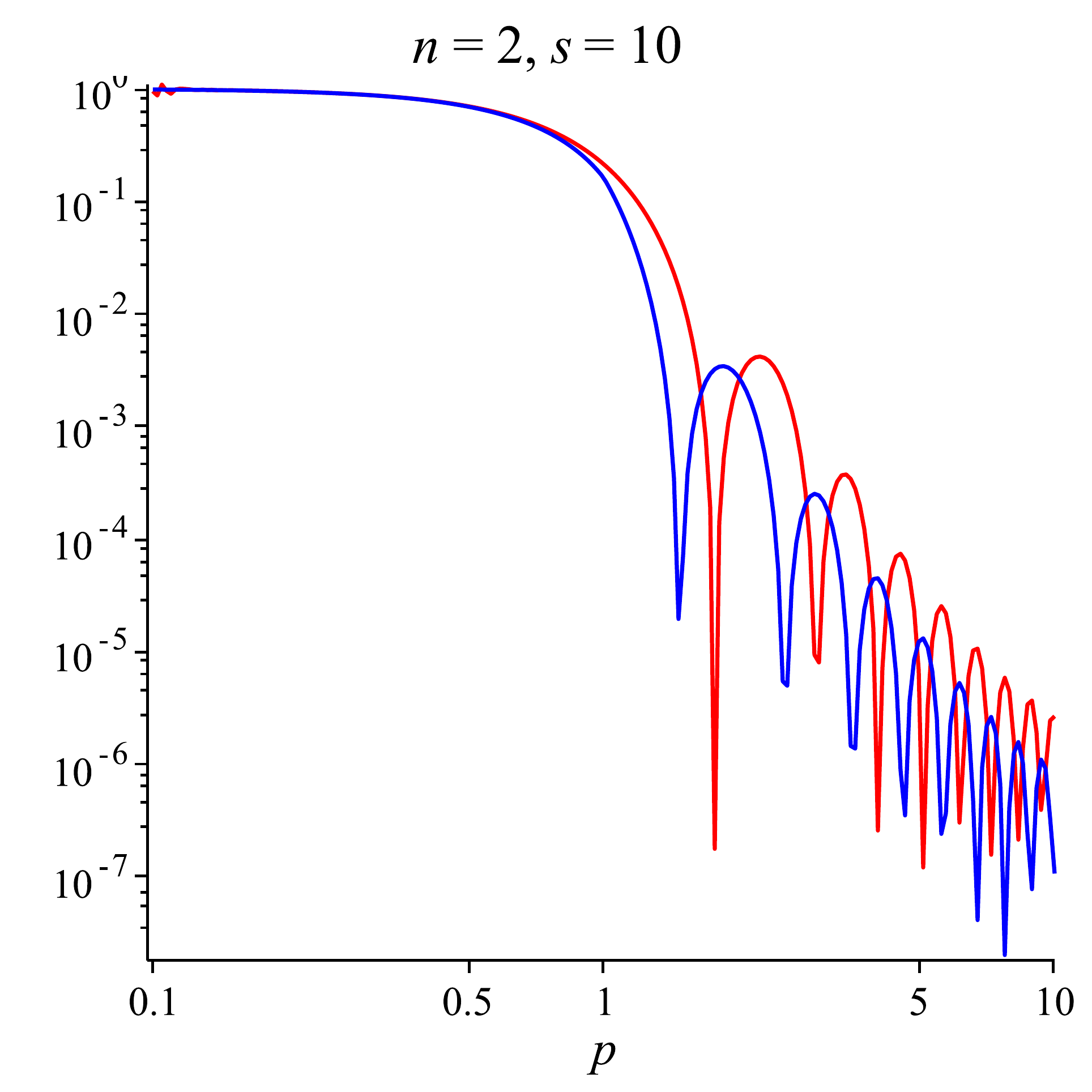}
\caption{Comparison of the power spectrum for the primordial tensor perturbation between the approximate solution \eqref{3in1} and exact one \eqref{Ph3exact} with different values of $n$ and $s$. Since the oscillation behavior of the power spectrum for the approximate solution appears only in the range of $k<k_2=sk_1$, in the figures, we only plot the range $p<s$ for different values of $s$ for comparison. As discussed below \eqref{Ph3longWKBser}, for long wavelengh modes with $k<k_1$, at $k^2$ order, the approximate solution has exactly the same form of exact solution. Hence, at range $k<k_1$, the blue line is consistent with the red one. When $k$ grows larger than $k_1$, the approximate solution begins to deviate from the exact one.This coincides with our expectation that WKB approximation is valid for small wavenumber. Upper: $n=1$ case which corresponds to a radiation-dominated intermediate stage with $w=1/3$. Down: $n=2$ case which corresponds to a matter-dominated intermediate stage with $w=0$.}\label{fig:compare}
\end{figure}

\subsection{Approximate power spectrum: summary}

Let us summarize the results we obtained with an approximate method.
The final power spectrum is expressed as
\be\label{3in1}
\mathcal P_h^{\mathrm{app}}=\left\{
\begin{matrix}
\displaystyle \mathcal{P}_{0}|\widetilde{F}|^2, & \displaystyle\text{for}~k<k_1\\
\\
\displaystyle \mathcal{P}_{0}\left|F\right|^2, & \;\;\;\;\displaystyle\text{for}~k_1<k<k_2\\
\\
\displaystyle \mathcal{P}_{0}s^{-2(n+1)}, & \displaystyle\text{for}~k>k_2
\end{matrix}
\right.
\ee
where $\widetilde{F}$ and $F$ are given by  \eqref{Flongexp} and \eqref{Fexp}, respectively. 

In order to compare with the power spectrum derived by the exact solution, 
in Fig.\ref{fig:compare}, 
we plot $P_h^{\mathrm{app}}$ and $P_h^{\mathrm{exact}}$ for $n=1$ and $n=2$ cases.\footnote{Note that we set $\tilde{z}_0=1$ and $y_0=1$ for numerical calculations.}.
Since the oscillation behavior appears only in the range of $k<k_2=sk_1$,
studies of which are our primary purpose, 
we only plot the range $p<s$ for several different values of $s$.
As we see from Fig.~\ref{fig:compare},  the approximation we employed can recover the
essential features of the power spectrum very well despite its crudeness.
In particular, the agreement of the approximate spectrum with the exact one is
surprisingly good on the long wavelength part at $k\leq k_1$ ($p\leq 1$).
This confirms our argument above \eqref{Ph3exactk2} that the suppression
is due to the correction term of $O(k^2)$ to the constant solution on superhorizon scales.

\section{Conclusion}\label{conclude}
In this paper, we have studied a double-inflationary model with an intermediate stage of 
decelerated expansion between the two stages of inflation. 
Such a scenario can be realized in a class of multi-field inflationary models
 where the inflaton at the first stage undergoes damped oscillations
while the potential is still non-vanishing at the minimum which leads to the second stage of inflation.
In such a model, it has been known that the scalar curvature perturbation exhibits
an oscillatory behavior with the amplitude substantially enhanced relative to that
at the first stage of inflation. Thus our primary purpose was to see if the
similar enhancement could appear also for the tensor perturbation.\\

  Assuming a general background which describes the double-inflationary scenario with
an intermediate break stage, the tensor power spectrum is 
derived in \eqref{Ph3exact} and \eqref{3in1} by the exact and approximate solutions, 
respectively. We have proved that although oscillations appear as a result of the 
presence of the break stage, the amplitude can never exceeds that at the first stage,
for any EOS parameter $w>-1/3$, as explicitly proved in Appendix~\ref{nogo}.
It seems that the main reason is the difference in the EOMs:
the scalar EOM contains pressure $p=w\rho$ and its first derivative explicitly that can 
undergo a first-order phase transition, while the tensor EOM depends only on the scale factor
and its first derivative which must be continuous.\\

Another finding is that there is an appreciable suppression in the amplitude
of the long wavelength modes, which never re-enter the horizon 
after the horizon exit during the first inflationary stage.
Naively one might expect that it would be frozen after horizon exist
so that the amplitude of the power-spectrum in the corresponding range 
would remain approximately the same as the original one given by the vacuum amplitude.
However, as seen from Fig.~\ref{fig:exact}, the suppression is unexpectedly strong.
For example, the amplitude at the critical wavelength that just touches the horizon scale
when the second stage of inflation starts may be suppressed by a factor of two.
This behavior is caused by the combined effect of 
the correction term of $O(k^2)$ to the constant solution
on superhorizon scales and the leading order decaying solution.
We have justified this argument by using approximate solutions
that contain both the correction term and the decaying modes. 
In fact, expanding the exact analytic power spectrum up through $O(k^2)$,
and comparing it with the one derived by the approximate method,
we find an exact agreement between the two.\\

We note that although we have not specified a detailed model of the double inflationary scenario,
it is highly probable that the existence of an intermediate break stage during inflation
will produce a curvature perturbation spectrum with pronounced features
which would be observationally ruled out if the features appear on the CMB scale,
as in the model studied in~\cite{Polarski:1994bk}. 
Thus the scale that exits the horizon during the intermediate stage must be either 
unobservably large or much smaller than the CMB scale where we have 
virtually no constraint on the shape of the curvature perturbation spectrum. In fact, 
as in a double inflationary model discussed in~\cite{Pi:2018}, a prominent peak 
 in the spectrum on a very small scale may result in a copious
 production of primordial black holes which may have interesting cosmological implications.
\\

In this work, in order to find the analytic solutions, we have assumed the exact deSitter 
background for the double inflationary stages and a constant EOS during the break stage. 
It would be the next step to consider the case
where the background of double inflation is described by quasi-deSitter, while the EOS during the break stage
evolves. There are several possible models to realize 
this scenario: one possible case is to enhance the value for parameter $\mu$, or include
the higher order terms in the potential of the $\chi$ field in Ref.~\cite{Pi:2018}. Another case could
be the $\alpha$-attractor-type multiple inflation model (e.g. see~\cite{Maeda:2018sje}). The
investigation of the detailed models will be a future work.

\vspace{5mm}
\noindent {\bf Acknowledgments}

We thank Shinji Tsujikawa for useful discussions.  
SP and MS were supported by the MEXT/JSPS KAKENHI Nos. 15H05888 and 15K21733, and by the World Premier International Research Center Initiative (WPI Initiative), MEXT, Japan.
YZ was supported by the NSFC grant No. 11605228, 11673025, 11720101004, by the MEXT Grant-in-Aid for Scientific Research on Innovative Areas No.15H05888, and by JSPS Grant-in-Aid for Young Scientists (B) No.15K17632.

\appendix
\section{Deduction of the form of scale factor}\label{app:a}
In this appendix we deduce expressions for the scale factor in eq.~(\ref{a-eta}).
The Friedman equation reads as
\begin{align}\label{Fried}
\mathcal{H}^2=\frac{\kappa^2}{3}a^2\rho\,,
\end{align}
where $\mathcal{H}\equiv a'/a$. Taking derivative with respect to conformal time $\eta$ on both sides, we have
\begin{align}
2\mathcal{H}\mathcal{H}'&=\frac{\kappa^2}{3}(2a'a\rho+a^2\rho')\nonumber\\
&=\frac{\kappa^2}{3}\left(\frac{6}{\kappa^2}\mathcal{H}^3+a^2\rho'\right)\,,
\end{align}
from which we obtain
\begin{align}\label{Frieddr}
a^2\rho'=\frac{6\mathcal{H}}{\kappa^2}\left(\mathcal{H}'-\mathcal{H}^2\right)\,.
\end{align}

On the other hand, the continuity equation reads as:
\begin{align}\label{Conti}
\rho'+3\mathcal{H}\rho(1+w)=0\,.
\end{align}
Inserting Eqs.~(\ref{Fried}) and (\ref{Frieddr}) into (\ref{Conti}), we have
\begin{align}
\frac{\mathcal{H}'}{\mathcal{H}^2}=-\frac{1+3w}{2}\,.
\end{align}
This equation can be integrated once to obtain
\begin{align}\label{Heq}
\mathcal{H}=2g(\eta)\,,
\end{align}
where we define the function
\begin{align}
g(\eta)\equiv\left[\int_{\eta_0}^{\eta}(1+3w)d\eta'\right]^{-1}\,.
\end{align}
Eq.~(\ref{Heq}) can be further integrated once to obtain
\begin{align}
a(\eta)=\exp\left(2\int_{\tilde{\eta}}^{\eta}g(\eta')d\eta'\right)\,.
\end{align}
where $\eta_0$ and $\tilde{\eta}$ are integration boundaries. In case of $w=const$, the scale factor can be expressed as
\begin{align}\label{afunction}
a(\eta)=a_0\left(\eta-\eta_0\right)^{\frac{2}{1+3w}}\,,
\end{align}
where $a_0\equiv\left(\tilde{\eta}-\eta_0\right)^{-\frac{2}{1+3w}}$. Comparing \eqref{afunction} with \eqref{a-eta}, we have $n=2/(1+3w)$.

\section{Expressions for integration constants of scale factor}\label{inteconstant}
In this appendix, we show the details for the deduction of integration constants $\gamma_1$, $\gamma_2$, $\alpha$ and $\beta$ by matching the solutions for scale factor at boundaries $\eta_1$ and $\eta_2$. Firstly, we multiply the first line of \eqref{a-eta} with $n/(\eta_1-\beta)$, we have 
\be
-\frac{n}{H_1(\eta_1-\gamma_1)(\eta_1-\beta)}=n\alpha(\eta_1-\beta)^{n-1}=\frac1{H_1(\eta_1-\gamma_1)^2},
\ee
which gives
\be\label{beta1}
\beta=(n+1)\eta_1-n\gamma_1.
\ee
Similarly from the third line of \eqref{a-eta} we have
\be\label{beta2}
\beta=(n+1)\eta_2-n\gamma_2.
\ee
Then we have
\be\label{c2-c1}
\gamma_2=\frac{n+1}{n}(\eta_2-\eta_1)+\gamma_1.
\ee
Next, we substitute \eqref{beta1} into the first line of \eqref{a-eta} to obtain
\be\label{alpha1}
\frac1{\alpha H_1}=n^n(\gamma_1-\eta_1)^{n+1}.
\ee
Similarly we have
\be\label{alpha2}
\frac1{\alpha H_2}=n^n(\gamma_2-\eta_2)^{n+1}.
\ee
Then we have
\be\label{def:s}
\frac{\gamma_2-\eta_2}{\gamma_1-\eta_1}=\left(\frac{H_1}{H_2}\right)^{1/(n+1)}\equiv s.
\ee
Then,~\eqref{def:s} together with \eqref{c2-c1} gives the solution
\bea\label{result:c1}
\gamma_1&=&\eta_1+\frac{\eta_2-\eta_1}{n(s-1)},\\\label{result:c2}
\gamma_2&=&\eta_2+\frac{s(\eta_2-\eta_1)}{n(s-1)}.
\eea
We see from the definition \eqref{def:s} that $s\equiv(H_1/H_2)^{1/(n+1)}>1$, which guarantees the trivial fact that $a(\eta_2)>a(\eta_1)$. Then, substituting \eqref{result:c1} into \eqref{alpha1} and \eqref{beta1}, we have
\bea\label{result:alpha}
\alpha&=&\frac{n}{H_1}\left(\frac{s-1}{\eta_2-\eta_1}\right)^{n+1},\\\label{result:beta}
\beta&=&\eta_1-\frac{\eta_2-\eta_1}{s-1}.
\eea

\section{Detailed deduction of the form of exact solutions}\label{Normalization}
In this appendix, we show the detailed deduction of the expression for the exact solution $h_{\mathrm{II}}$. After $\eta>\eta_1$, the scale factor changes according to \eqref{a-eta}. Defining $v_{\mathrm{II}}\equiv a_{\mathrm{II}}h_{\mathrm{II}}\mpl /2$, the EOM for the break stage reads as
\begin{align}
v_{\mathrm{II}}''+\left[k^2-\frac{n(n-1)}{(\eta-\beta)^2}\right]v_{\mathrm{II}}=0\,,
\end{align}
This equation has a typical form of Bessel equation. Hence, it can be solved as
\begin{align}\label{v2exp}
v_{\mathrm{II}}&=\sqrt{\frac{z_2}{k}}\left[D_1J_{\nu}(z_2)+D_2Y_{\nu}(z_2)\right]\nonumber\\
&=\frac{1}{2}\sqrt{\frac{z_2}{k}}\left[\left(D_1-iD_2\right)H_{\nu}^{(1)}(z_2)+\left(D_1+iD_2\right)H_{\nu}^{(2)}(z_2)\right]\,,
\end{align}
where $H_{\nu}^{(1)}(z)$ and $H_{\nu}^{(2)}(z)$ are the Hankel functions of first and second kind, respectively. $D_1$ and $D_2$ are two coefficients which are functions of wavenumber $k$. Using the relation $H^{(1) *}_{\nu}(z)=H^{(2)}_{\nu}(z)$, we have
\begin{align}
v_{\mathrm{II}} v_{\mathrm{II}}^{*\prime}-v_{\mathrm{II}}^{\prime}v_{\mathrm{II}}^*&=z_2\left(H^{(1)}_{\nu}\frac{dH^{(2)}_{\nu}}{dz_2}-H^{(2)}_{\nu}\frac{dH^{(1)}_{\nu}}{dz_2}\right)\left(D_1D_2^*-D_1^*D_2\right)\nonumber\\
&=\frac{2}{\pi}\left(D_1D_2^*-D_1^*D_2\right)\,,
\end{align}
where we have used the relation $H^{(1)}_{\nu}~dH^{(2)}_{\nu}(z)/dz-H^{(2)}_{\nu}~dH^{(1)}_{\nu}/dz=-4i/(\pi z)$. Since we have the relation $v_{\mathrm{II}} v_{\mathrm{II}}^{*\prime}-v_{\mathrm{II}}^{\prime}v_{\mathrm{II}}^*=i$, then we obtain
\begin{align}\label{D12relation}
D_1D_2^*-D_1^*D_2=i\frac{\pi}{2}\,.
\end{align}
Therefore, we define the normalized coefficients $C_1$ and $C_2$ as
\begin{align}\label{normalC12}
C_1\equiv\sqrt{\frac{2}{\pi}}D_1,\qquad C_2\equiv i\sqrt{\frac{2}{\pi}}D_2\,,
\end{align}
which yields the normalization relation
\begin{align}\label{C12relation}
C_1C_2^*+C_1^*C_2=1\,.
\end{align}
Inserting \eqref{normalC12} into \eqref{v2exp}, using the definition $v_{\mathrm{II}}\equiv a_{\mathrm{II}}h_{\mathrm{II}}\mpl /2$, we obtain the normalized form for $h_{\mathrm{II}}$ as
\begin{align}\label{h2exactapp}
h_{\mathrm{II}}&=\frac{2}{a_{\mathrm{II}}\mpl}\sqrt{\frac{z_2}{k}}\left[D_1J_{\nu}(z_2)+D_2Y_{\nu}(z_2)\right]\nonumber\\
&=\frac{\sqrt{2\pi}}{\alpha\mpl}\left(\frac{z_2}{k}\right)^{\frac{1}{2}-n}\left[C_1J_{\nu}(z_2)-iC_2Y_{\nu}(z_2)\right]\,.
\end{align}
The coefficients $C_1$ and $C_2$ will be determined by matching $h_{\mathrm{I}}$ and $h_{\mathrm{II}}$ at point $\eta_1$.

\section{The approximate solutions with suppression term outside the horizon}\label{WKBsuppress}
Let us consider the range $\eta_1<\eta<\eta_k^{(2)}$ which corresponds to the break stage with $a\propto(\eta-\beta)^n$. As argued in Section.\ref{WKBsec}, the constant super-horizon solution \eqref{h>} is obtained in the limit where $k/(aH)\rightarrow0$, or equivalently, the $k^2h$ term in \eqref{basicEOM} is simply neglected. Hence, the correction term of the lowest order would appear once we take into account the $k^2h_2$ term, which implies that the correction term of the lowest order should be proportional to $z_2^2$. Therefore, we assume that $h_2^{(\mathrm{I})}$ including the correction term of the lowest order can be expressed in the following form
\be\label{h2supp}
h_2^{(\mathrm{I})}=C+\widetilde{C}z_2^2\,,
\ee
where $C$ and $\widetilde{C}$ are two constants.  The constant $\widetilde{C}$ can be determined when we insert \eqref{h2supp} into the EOM
\be\label{h2EOM}
h_2^{(\mathrm{I})\prime\prime}+\frac{2n}{\eta-\beta}h_2^{(\mathrm{I})\prime}+k^2h_2^{(\mathrm{I})}=0,
\ee
and expand this equation to the order $k^2$. For the consistency of the equation, we obtain
\be\label{Dexpress}
\widetilde{C}=-\frac{C}{2(1+2n)}\,.
\ee

Thus, taking into account the lowest order suppression term, together with the decaying mode, the amplitude of tensor perturbations outside the horizon can be expressed as
\be\label{h2sup}
h_2^{(\mathrm{I})}=C\left[1-\frac{z_2^2}{2(1+2n)}\right]+Dz_2^{1-2n}\,.
\ee
The constants $C$ and $D$ will be determined from the matching conditions. For deSitter background $n=-1$, we have
\be\label{h1sup}
h_1^{(\mathrm{II})}=A\left(1+\frac{z_1^2}{2}\right)+Bz_1^{3}\,.
\ee

\section{Proof of no-go theorem of the enhancement of the power spectrum for primordial tensor perturbations}\label{nogo}

For simplicity of symbols, let us rewrite the expression \eqref{Ph3longWKBser} as follows:
\begin{align}\label{PWKBdef}
\frac{\mathcal{P}_{k<k_1}}{\mathcal{P}_{0}}=1-\frac{(2\nu+3)L}{48\nu(1+\nu)}\left(\frac{p}{s}\right)^2+\mathcal{O}\left(p^4\right)\,,
\end{align}
where
\begin{align}\label{Lexp}
L\equiv s^{2}\nu(5+6\nu)-3(2\nu-1)(\nu+1)-s^{-2\nu}(2\nu+3)\,.
\end{align}
Since we are considering an expanding universe, the parameter $s\equiv (H_1/H_2)^{1/(n+1)}>1$. According to the signature of $\nu$, we divide the discussion of the signature of $L$ into three cases as follows.

\subsection{$\nu>0$ case}
In this case, we have
\begin{align}
L>\nu(5+6\nu)-3(2\nu-1)(\nu+1)-(2\nu+3)=0\,,
\end{align}
which leads to the correction term
\begin{align}\label{case1}
-\frac{(2\nu+3)L}{48\nu(1+\nu)}\left(\frac{p}{s}\right)^2<0\,.
\end{align}

\subsection{$-1/2<\nu<0$ case}
Similarly as $\nu>0$ case, here we have
\begin{align}
L<\nu(5+6\nu)-3(2\nu-1)(\nu+1)-(2\nu+3)]=0\,,
\end{align}
so we arrive at the same conclusion as shown in \eqref{case1}.

\subsection{$\nu=0$ case}
In this case, we expand $L$ around $\nu\ll1$ so that
\begin{align}
L=3-3\exp\left(-2\nu\ln s\right)=6\nu\ln s+\mathcal{O}(\nu^2)\,.
\end{align}
Hence, the correction term
\begin{align}\label{case3}
-\frac{(2\nu+3)L}{48\nu(1+\nu)}\left(\frac{p}{s}\right)^2=-\frac{(2\nu+3)\ln s}{8(1+\nu)}\left(\frac{p}{s}\right)^2<0\,.
\end{align}

Therefore, the correction of order $k^2$ is always negative. Hence, we reach at the conclusion: provided that the background of the break stage (between the two inflationary stages) describes an expanding universe, the corresponding power spectrum of the primordial tensor perturbations is always suppressed for long wave-length modes. 

%
%
%

\end{document}